\documentclass[graybox, nosecnum]{svmult}

\usepackage{mathptmx}       
\usepackage{helvet}         
\usepackage{courier}        
\usepackage{type1cm}        
\usepackage{makeidx}         
\usepackage{graphicx}        

\usepackage{tabularx}\usepackage{multicol}        
\usepackage{multirow}
\usepackage{makecell}
\usepackage[bottom]{footmisc}
\usepackage{hyperref}        
\usepackage{soul}            
\hypersetup{colorlinks=true,urlcolor=blue}
\usepackage{amsmath,amssymb}
\usepackage{tikz}
\usetikzlibrary{calc,patterns,angles,quotes}
\usepackage{siunitx}
\usepackage{multirow}
\usepackage {threeparttable}
\usepackage{comment}
\usepackage[square,numbers]{natbib}

\newcommand*{\myalign}[2]{\multicolumn{1}{#1}{#2}}
\makeindex             

\begin{document}
\title*{How to detect Gamma-rays from ground: an introduction to the detection concepts}
\author{Manel Errando and Takayuki Saito}
\institute{Manel Errando \at Department of Physics, Washington University in St Louis, St Louis, MO 63110, USA;  \email{errando@wustl.edu}
\and Takayuki Saito \at Institute for Cosmic Ray Research, The University of Tokyo, Kashiwa, 277-8582 Chiba, Japan;  \email{tsaito@icrr.u-tokyo.ac.jp}}

\maketitle

\abstract{
Indirect detection of gamma rays with ground-based observatories is currently the most sensitive experimental approach to characterize the gamma-ray sky at energies $>0.1$\,TeV. Ground-based detection of gamma-rays relies on the electromagnetic showers that gamma rays initiate in the Earth's atmosphere. In this chapter we will review the properties of electromagnetic air showers as well as the differences with respect to cosmic-ray showers that enable the rejection of the cosmic ray background. The experimental techniques that have been developed for ground-based detection of gamma  rays will be introduced. These fall onto three main categories: air shower particle detectors, sampling Cherenkov arrays, and imaging atmospheric Cherenkov telescopes.  Hybrid concepts as well as other experimental approaches are also discussed. 
}
\section{Keywords} 
Gamma rays, TeV, Cherenkov, Air showers, cosmic rays, ground-based observatories
\section{Introduction}

The Earth's atmosphere is opaque to gamma rays. Cosmic gamma rays interact with the upper atmosphere producing showers of energetic electrons, positrons and photons. The detection and characterization of these electromagnetic showers enables the study of cosmic gamma-ray sources from the ground. 

Current-generation ground-based observatories achieve their peak sensitivity for gamma rays in the TeV energy scale. The gamma-ray flux from the Crab Nebula above 1 TeV is $\sim 2 \times 10^{-7}\,\textrm{m}^{-2}\,\textrm{s}^{-1}$. TeV observatories need to realize effective areas $>10^{4}\,\textrm{m}^2$ to detect a few photons from the Crab Nebula in one hour of exposure. Such large instruments cannot be mounted on an orbital platform and must be installed on the Earth's surface, making the Earth's atmosphere an integral part of any ground-based gamma-ray observatory. Rather than detecting the passage of primary gamma rays through an instrument, ground-based detection is achieved in an indirect fashion by recording the secondary particles and radiation produced by the interaction of the primary gamma-ray with the atmosphere. These secondaries comprise what is known as extensive air showers.

There are two main techniques to detect cosmic gamma rays from the ground. The first measures the passage of the secondary charged particles that make the extensive air shower through a surface detector array. The direction of the primary gamma ray is determined by arrival time differences recorded at different detectors as the shower front sweeps through the detector plane. The second technique uses optical telescopes to focus Cherenkov light produced by shower particles onto photon detectors. The first generation of Cherenkov telescopes had a single photomultiplier seeing each mirror. The second generation used the imaging technique with multi-pixel photomultiplier cameras and exploited results from Monte Carlo simulations of electromagnetic and hadronic air showers to improve the angular resolution and background rejection. 
This led to the detection of the Crab Nebula by the Whipple 10\,m observatory in 1989 \cite{1989ApJ...342..379W}, the first significant detection of an astrophysical gamma-ray source with a ground-based observatory. Three decades later, the LHAASO reported the detection of a population of galactic PeV sources using a particle sampling array. Many observatories have been in operation during this time, including sampling and imaging Cherenkov telescopes as well as particle sampling arrays, leading to the development of ground-based gamma-ray astronomy \cite{2009ARA&A..47..523H}.

This chapter discusses the development of electromagnetic and hadronic showers in the Earth's atmosphere and describes the operating principles of the different types of ground-based gamma-ray observatories.  
For a more comprehensive discussion on electromagnetic and hadronic showers, along with a review of the basic physics processes that influence their development, the reader is directed to the book by Gaisser, Engel \& Resconi \cite{2016crpp.book.....G}. A classic discussion of electromagnetic air showers that includes parameterizations that are still used to this day can be found in the review by Rossi \& Greisen \cite{1941RvMP...13..240R}. Finally, a more compact description of shower physics along with the principles of ground-based detection of gamma rays is given in the review by Aharonian, Buckley, Kifune \& Sinnis \cite{2008RPPh...71i6901A}.

\section{Electromagnetic air showers}
\label{sec:em_showers}
\begin{figure}[tb]
\centering
 \includegraphics[height=8cm]{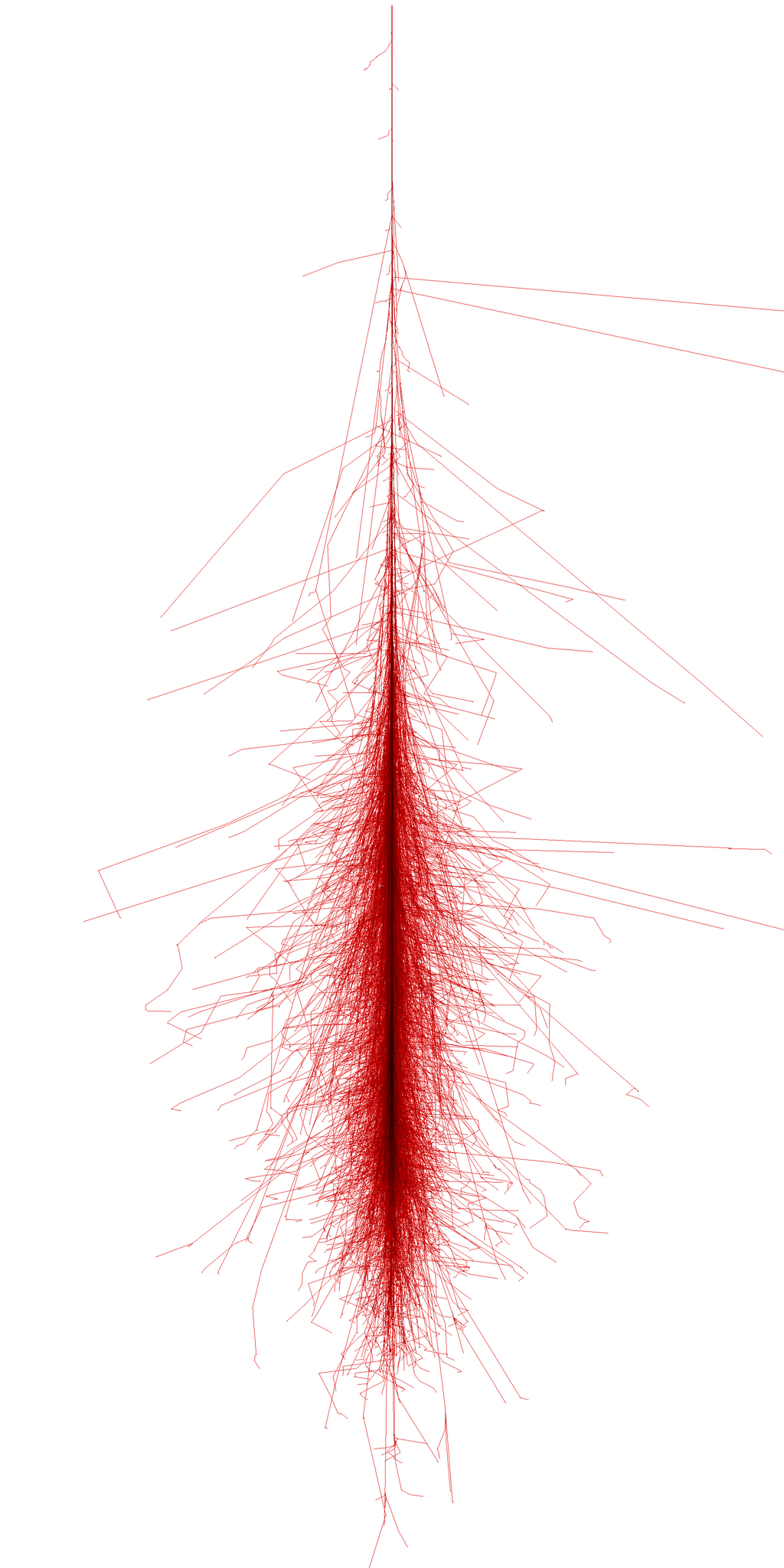}
 \hspace{3cm}
  \includegraphics[height=8cm]{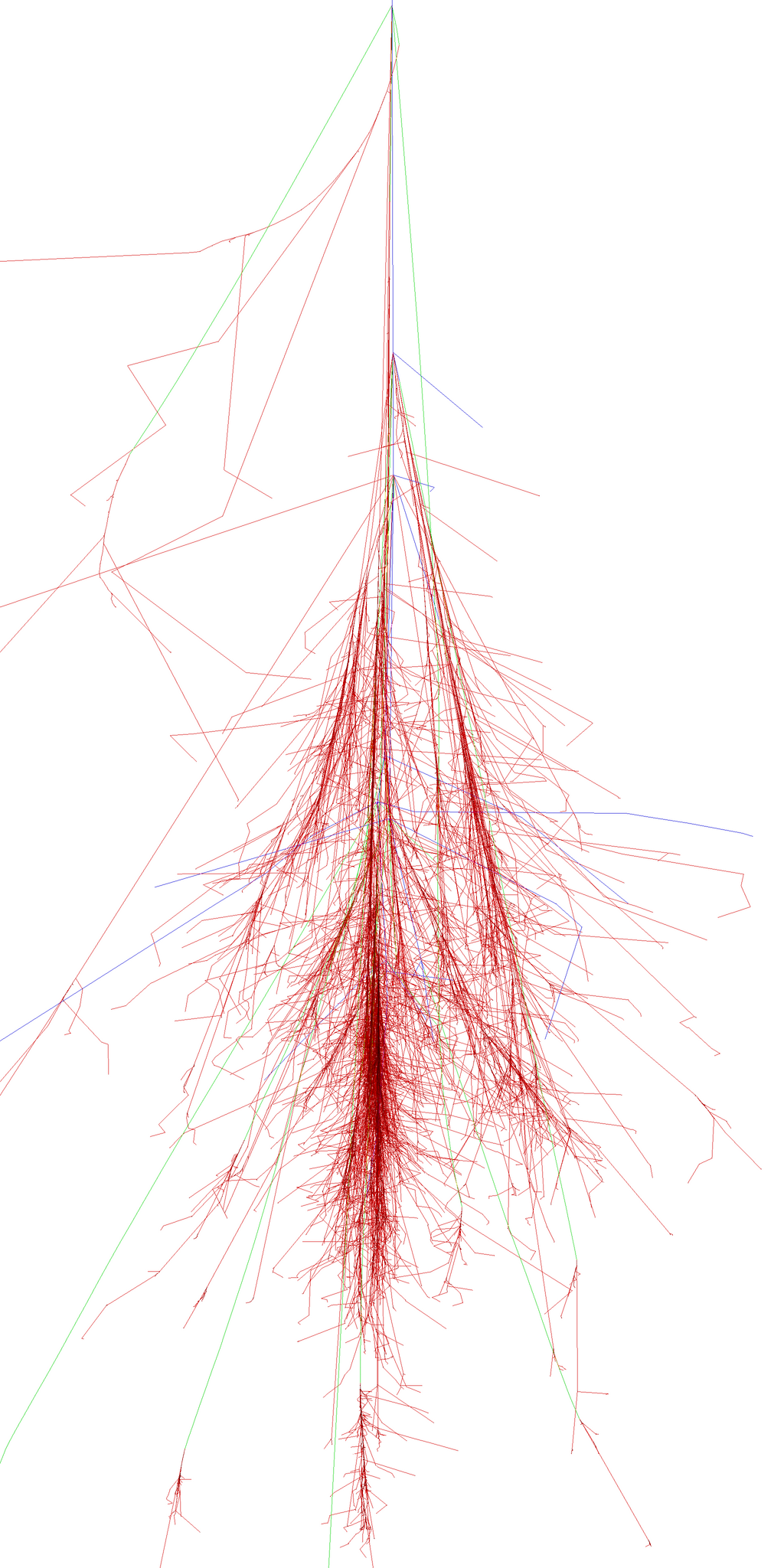}
  \put(-255,220){\colorbox{white}{\color{black} \bf 0.1\,TeV photon}}
\put(-120,220){\colorbox{white}{\color{black} \bf 0.1\,TeV proton}}
     \caption{Simulated showers initiated by a 100\,GeV gamma ray (\textit{left}) and a 100\,GeV proton (\textit{right}) in the Earth's atmosphere. Red tracks indicate secondary electrons, positrons, and photons with energy $E>0.1$\,MeV. Green and blue tracks show muons and hadrons with $E>0.1$\,GeV. The first interaction is fixed at 30\,km height and the lateral scale is $\pm 5$\,km. Adapted from \cite{knapp}.}
        \label{f:showers}
\end{figure}

Gamma-ray photons incident in the Earth's atmosphere will pair produce in the presence of the Coulomb field of an atmospheric nucleus. The resulting electron-positron pair will subsequently produce multiple generations of secondary photons and pairs via bremsstrahlung and pair production, leading to the development of an electromagnetic cascade (Fig.~\ref{f:showers}). Eventually, the energy is dissipated by ionization of the medium (the Earth's atmosphere) by all the electrons and positrons in the cascade. Photon or cosmic-ray-initiated cascades in the Earth's atmosphere are often referred to as air showers.  

In this section we will provide some analytical approximations that will give the reader a quantitative understanding of the aspects of electromagnetic showers that are most relevant to the indirect detection of gamma rays from the ground.

The three processes that govern the development of electromagnetic particle showers initiated by gamma rays are bremsstrahlung, pair production, and ionization losses. Bremsstrahlung energy losses for electrons and positrons can be characterized by a radiation length $X_0$ that depends on the medium and is defined as the amount of matter a high-energy electron has to traverse to lose all but $1/e$ of its energy. 
The processes of pair production and bremsstrahlung are very closely related. Their Feynman diagrams are variants of one another. As a result, the radiation length for bremsstrahlung is $7/9$ of the mean free path for pair production. Given the similarity of length scales between the two main processes driving the development of electromagnetic air showers, one can derive a simple analytical approximation for the development of the shower in which every $\ln (2)\,X_0$ every particle (photon, electron, or positron) produces two more particles that share its energy \cite{1954qtr..book.....H}. Eventually, the energy of the electrons and positrons reaches a critical energy $E_\textrm{c}$ where ionization takes over from bremsstrahlung as dominant energy loss mechanism for charged leptons and the electromagnetic shower dies out. In air, $E_\textrm{c}=88$\,MeV for electrons and 86\,MeV for positrons \cite{2020PTEP.2020h3C01P}. All particles in the shower are strongly collimated along the incident direction of the primary gamma ray that defines the shower axis. Multiple Coulomb scattering and, at second order, the deflection of charged particle trajectories by the Earth's magnetic field contribute to broadening the shower profile. Fig.~\ref{f:showers} shows the simulated particle tracks of an atmospheric air shower initiated by a gamma ray. 

\subsection{The Earth's atmosphere}
Particle showers develop on a medium (the Earth's atmosphere) that has increasing density as the shower progresses from high altitudes toward the ground level. The density profile of the atmosphere at mid-latitudes can be reasonably approximated by an exponential function
\begin{equation}
    \rho (z) = \rho_0 \, \exp(-z/H)
\end{equation}
where $z$ is the vertical height measured above sea level, $\rho_0 \sim 1.225 \times 10^{-3}$\,g\,cm$^{-3}$ and $H\sim 8.4$\,km it the atmospheric scale height. Most practical applications use tabulated atmospheric models based on atmospheric profiling data that describe the experimental sites where gamma-ray observatories are located. An example of the atmospheric properties at certain relevant altitudes for one such model is given in Table~\ref{t:atm}.
Temperature effects close to the Earth's surface produce seasonal changes in density of the order of 3--5\% at 10\,km height increasing to $\sim 15$\% at 15\,km.

\begin{table}[b]
\centering  
\begin{tabular}{S[table-format=3.0]S[table-format=2.1]S[table-format=1.2]ccS[table-format=2.1]S[table-format=3.1]S[table-format=4.0]S[table-format=5.0]}
\hline\hline
{height} & {vert. depth} & {density}  & {Ch. th.} & {Ch. ang.}& \multicolumn{4}{c}{$N(z)$ for $E_\gamma$/TeV}\\
{$z$ [km]} & {$x^\prime/X_0$} & {$\rho$ [$10^{-3}\,$g/cm$^{3}$]} & {[MeV]} & {$\theta_\textrm{Ch}$ [$^\circ$]} & {0.1} & {1.0} & {10} & {100}\\
\hline
20 & 1.52 & 0.088& 80 & 0.36    \\
10 & 7.25 & 0.42 & 37 & 0.79    \\
5  & 15.0 & 0.74 & 28 & 1.05 & 21 & 490 & 7800 & 92000 \\
3. & 19.5 & 0.91 & 25 & 1.17 & 3.0 & 110 & 2800 & 51000 \\
1.5& 23.6 & 1.06 & 23 & 1.26 & 0.4 & 21  & 740  & 19000 \\
0. & 28.2 & 1.23 & 21 & 1.36 & 0.04 & 2.6 & 120  & 4100  \\
\hline
\hline

\end{tabular}
\caption{\label{t:atm} Atmospheric parameters from the U.S. Standard Atmosphere model that are relevant for the development of air showers and the production of Cherenkov light. Columns indicate the vertical height measured from sea level, vertical atmospheric depth in radiation lengths, local density of the atmosphere, threshold for production of Cherenkov light and Cherenkov angle for electrons, and the average expected number of shower particles for showers initated by gamma rays with $E_\gamma= 0.1$\,TeV, 1\,TeV, 10\,TeV, and 100\,TeV. Data from \cite{atm76,2016crpp.book.....G}.}
\end{table}

\begin{figure}[t!]
\centering
\begin{tikzpicture}[scale=0.7]
    \draw[thick] (5,0) -- (-5, 0);
    \node [left] at (-5,0) {$z=0$};
    \node [right,below] at (5,0) {ground level};
    \draw[thick] (5,4) -- (-5, 4);
    \node [left] at (-5,4) {$x=0$};
    \node [right,above] at (5,4) {top of the  atmosphere};
    \draw[dotted, thick]   (5,2) -- (-5, 2);
    \draw[thick, ->] (-3,0) -- (-3,2) node[midway,left] {$z$};
    \draw[thick, ->] (-3,0) -- (0,2) node[midway,below] {$\ell$};
    \draw[thick, <->] (0,2) -- (3,4) node[midway,below] {$x$};
    \draw[thick, <->] (-3,2) -- (-3,4) node[midway,left] {$x^\prime$};
    \node [above right] at (-2.9,0.4) {$\theta$};
    \draw[thick] (-3,0.5) arc (90:34:0.5);
\end{tikzpicture}
\caption{\label{fig:atm} Definition of the variables used to describe the Earth's atmosphere. Adapted from \cite{2016crpp.book.....G}.}
\end{figure}
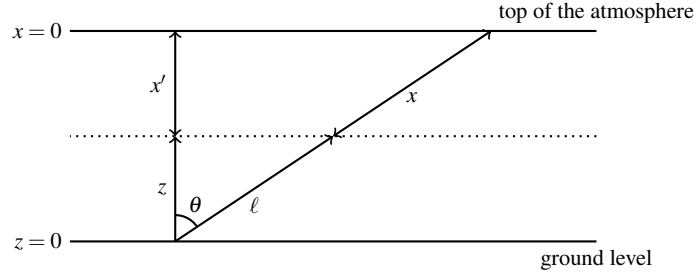

Let us define the total path length in the atmosphere of a particle with incident zenith angle $\theta$ moving on a straight trajectory from a vertical height $z$ to the ground as $\ell$ (Fig.~\ref{fig:atm}). For small zenith angles ($\theta \lesssim 65^\circ$) the curvature of the Earth can be neglected and $\ell = z / \cos \theta$.  
To remove the effect of the changing density of the atmosphere 
it is common to quote the depth $x$ of the particle track in units of g\,cm$^{-2}$ instead of the geometric path length $\ell$ and to scale it by the radiation length of the material, which is $X_0 = 36.6\,\mathrm{g}\,\mathrm{cm}^{-2}$ for dry air at 1\,atm of pressure \cite{2020PTEP.2020h3C01P}. The atmospheric depth of a particle entering the atmosphere and moving down to a height $z$ will then be
\begin{equation}
x = \int_\ell^\infty \rho (z = \ell \cos \theta)\, \textrm{d}\ell     
\end{equation}
which reduces to 
\begin{equation}
x^\prime = \int_{z}^\infty \rho (z^\prime)\, \textrm{d}z^\prime   
\end{equation}
for a vertically incident particle. In this framework, the atmosphere of the Earth can be considered an electromagnetic calorimeter with approximately 28 radiation lengths of low $Z$ material (Table~\ref{t:atm}).

\subsection{Longitudinal and lateral development of electromagnetic showers}
\label{sec:sh_devel}
\begin{figure}[!t]
\centering
 \includegraphics[height=8cm]{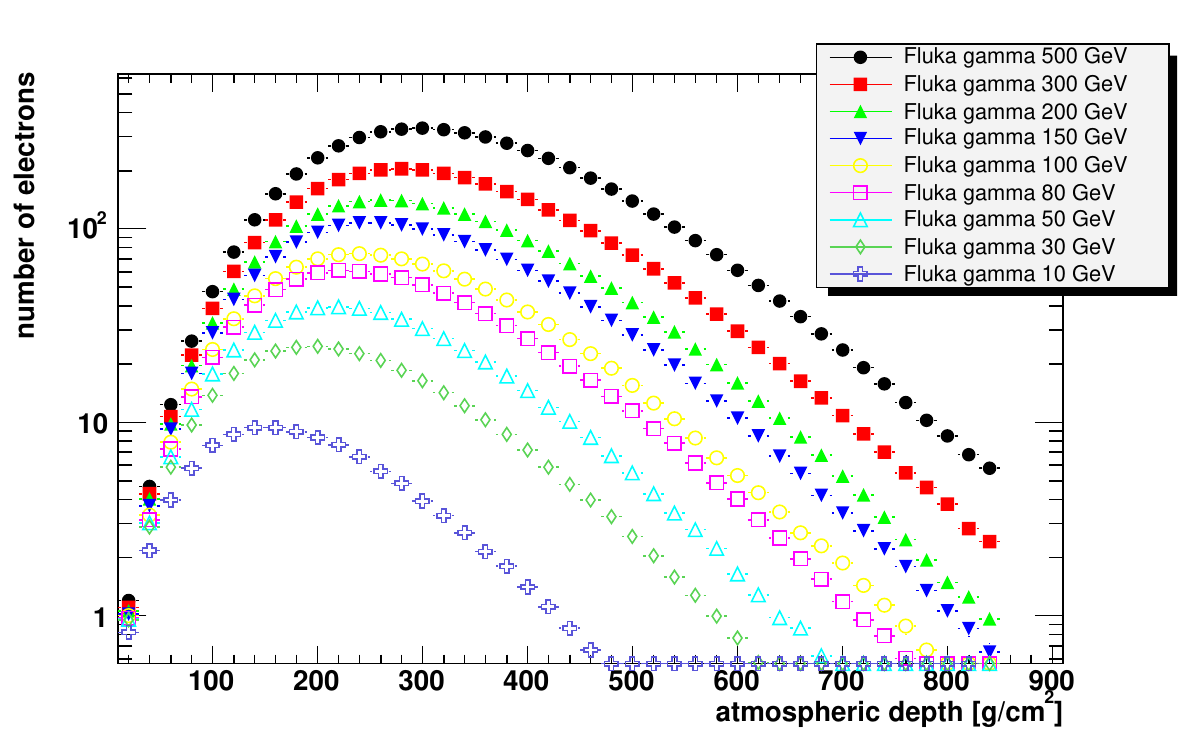}
      \caption{ Longitudinal development of electromagnetic air showers in the Earth's atmosphere. The average number of electrons as a function of atmospheric depth is shown for 1000 simulated vertical incidence showers with different primary energies. Simulations were performed using the \texttt{CORSIKA} package \cite{1998cmcc.book.....H}. Courtesy of Gernot Maier \cite{gernot_maier_2022_6037688}.}
        \label{f:lon_dist}
\end{figure}

A primary gamma ray with energy $E_\gamma$ will develop an electromagnetic shower in the atmosphere. At ground level, a typical shower front has a radius of 130\,m and a thickness of 1--2\,m at the shower core, growing wider toward the edges of the front. The number of electromagnetic particles in the shower as a function of atmospheric depth is shown in Fig.~\ref{f:lon_dist} and can be described by the analytical Approximation B approach from \cite{1941RvMP...13..240R}. After the first interaction, the number of electrons and positrons grows rapidly until reaching a  shower maximum, which will occur at an atmospheric depth
\begin{equation}
    x_\textrm{max} = X_0 \ln (E_\gamma/E_\textrm{c})
\end{equation}
For a 1\,TeV primary gamma ray the shower maximum occurs $\sim 10$\,km above sea level. After the shower maximum, the number of particles decreases by a factor of $\sim 1.65$ for each additional radiation length that is transversed. 

The number $N$ of secondary electrons and positrons at a given atmospheric depth $x$ can be approximated by \cite{1982JPhG....8.1461H}
\begin{equation}
    N(x) = \dfrac{0.31}{\sqrt{\ln (E_\gamma/E_\textrm{c})}} \exp \left[ \tfrac{x}{X_0} \left(1-1.5\ln{s}\right)\right]
\end{equation}
where $s$ is the shower age parameter that describes the stage of development of the shower and is given by
\begin{equation}
    s = \dfrac{3}{1+2\,\ln(E_\gamma/E_\textrm{c})/(x/X_0)}
\end{equation}
Table~\ref{t:atm} lists the average number of particles that reach different altitudes for gamma-ray showers with $E_\gamma = 0.1$\,TeV, 1\,TeV, 10\,TeV, and 100\,TeV. At the shower maximum, the total number of particles is approximately $N (x_\textrm{max})\sim10^{3}\,E_\gamma/\textrm{TeV}$.

\begin{figure}[!t]
\centering
 \includegraphics[width=0.99\textwidth, clip, trim={0 0 0 0cm}]{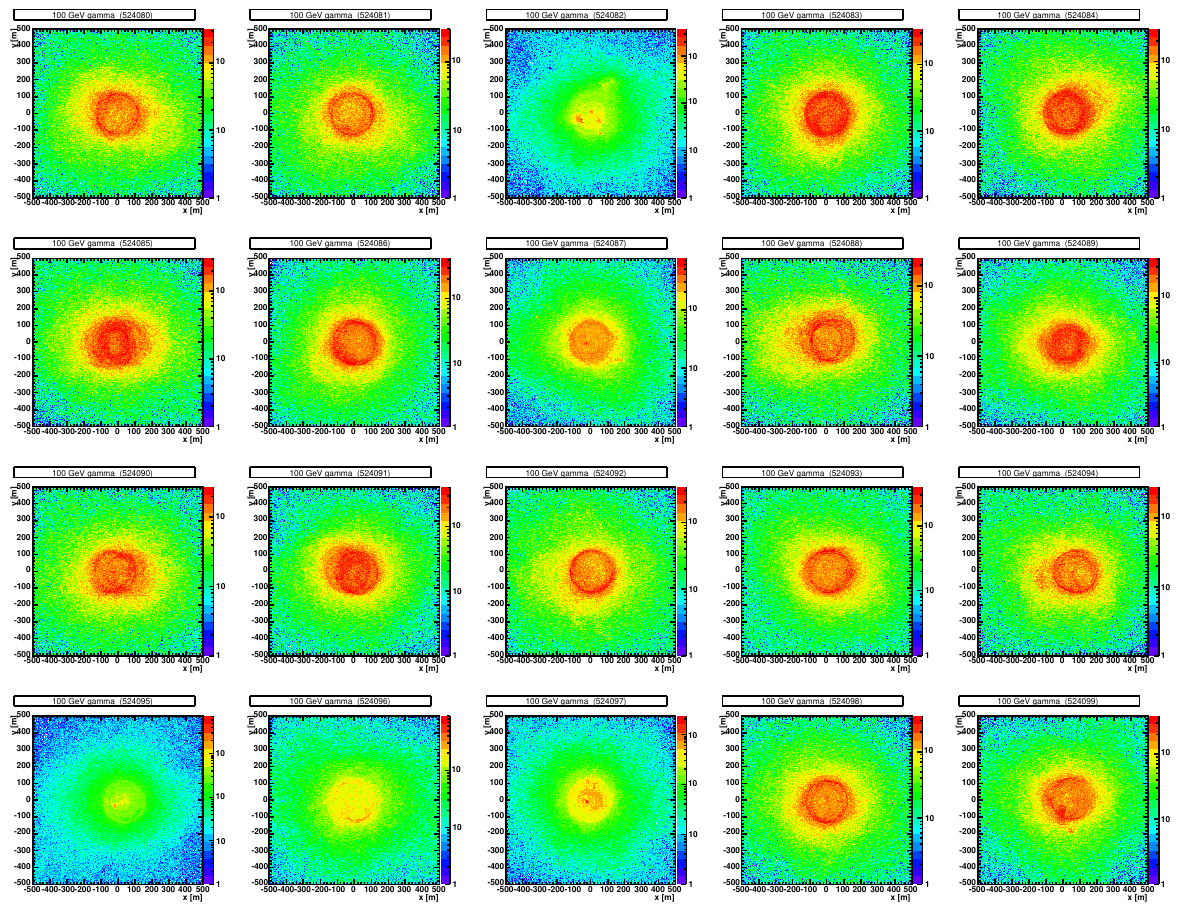}
      \caption{Effect of shower-to-shower variations in electromagnetic cascades. The figures show the density of Cherenkov photons hitting the ground for ten simulated 100\,GeV photon showers. Observatory altitude is set at 1,800\,m above sea level. 
      Simulations were performed using the \texttt{CORSIKA} package \cite{1998cmcc.book.....H}. Courtesy of Gernot Maier \cite{gernot_maier_2022_6037688}.}
        \label{f:ch_fluctuations}
\end{figure}

The leading factor responsible for shower-to-shower variance is fluctuations in the depth of the first interaction. The probability that a primary gamma ray will propagate to an atmospheric depth $x$ without interacting is $P(x) = \exp(-9x/7X_0)$. If we consider a collection of showers with the same $E_\gamma$ measured at the same atmospheric depth, the fluctuations in the number of secondary particles can be parameterized as \cite{2016crpp.book.....G}
\begin{equation}
    \delta \ln{N} \sim \dfrac{9}{14} \left(s - 1 - 3\ln{s}\right)
\end{equation}
Shower fluctuations are proportional to $N$ and are smallest if the shower is sampled at or close to the shower maximum ($s=1$), emphasizing the need for particle-sampling arrays to be located at high altitudes. The effect of shower-to-shower fluctuations can be visualized in Fig.~\ref{f:ch_fluctuations}. 

The lateral spread of electromagnetic showers determines the size of the particle pool on the ground that particle sampling arrays can use to detect gamma-ray showers. 
The size of the shower front can be characterized by the Moli\`{e}re radius, which is a property of the material in which the shower develops and can be expressed as 
\begin{equation}
    R_\textrm{M} = 0.24 \dfrac{X_0}{\rho}
\end{equation}
The Moli\`{e}re radius indicates the radius of the cylinder that contains 90\% of an electromagnetic shower in a given medium. At sea level $R_M \sim 80$\,m, but it increases with altitude as the atmosphere becomes less dense (see Table~\ref{t:atm}). The lateral spread of the shower is driven by Coulomb scattering of the low-energy secondaries close to the critical energy and is well characterized by $R_\textrm{M}$. Higher energy particles have their characteristic lateral spread reduced by a factor of $\sim E_\textrm{c}/E$. 

\begin{figure}[!t]
\centering
\raisebox{4.5cm}{\includegraphics[width=4.8cm, angle=270]{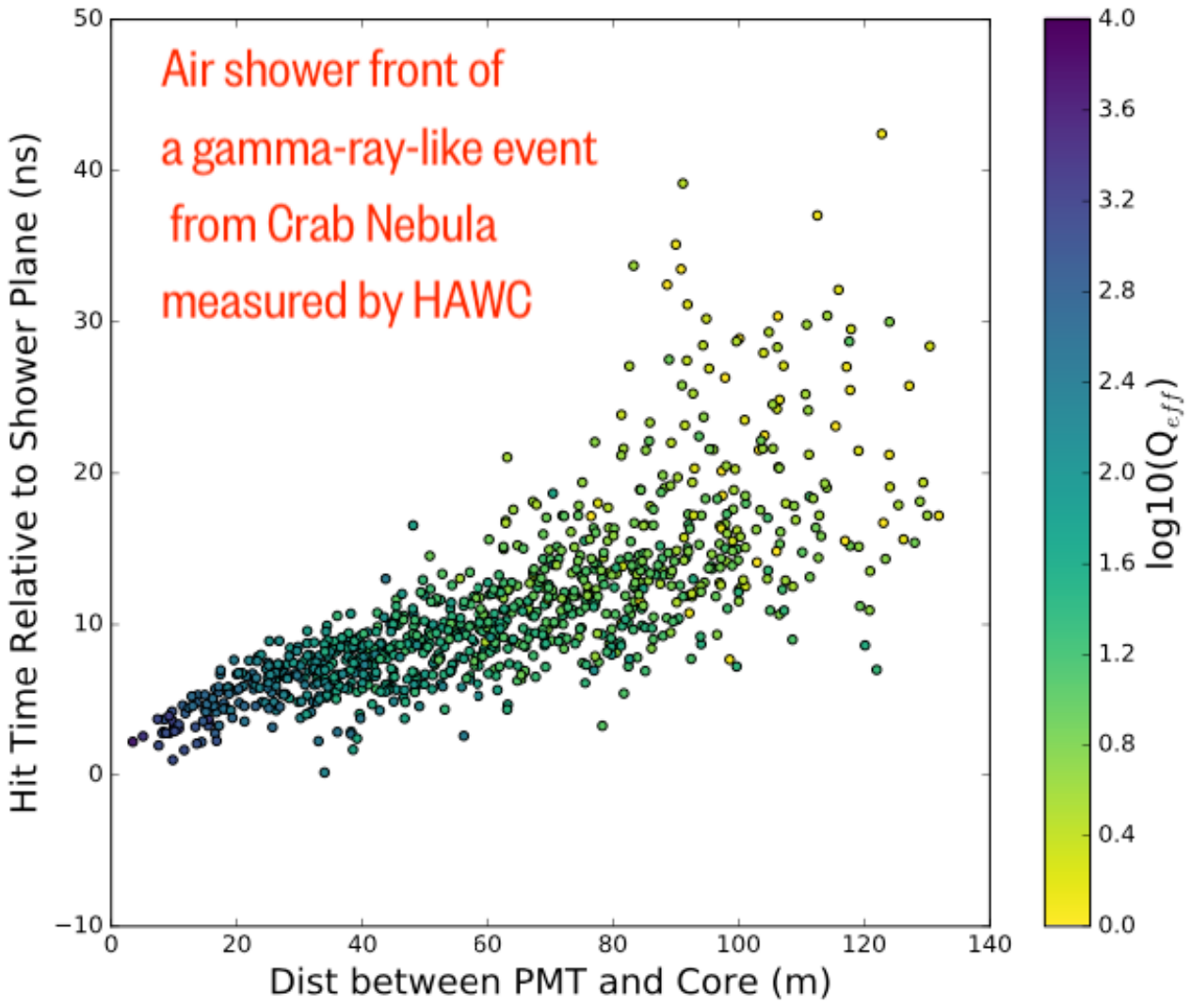}}
\hfill
 \includegraphics[height=4.3cm]{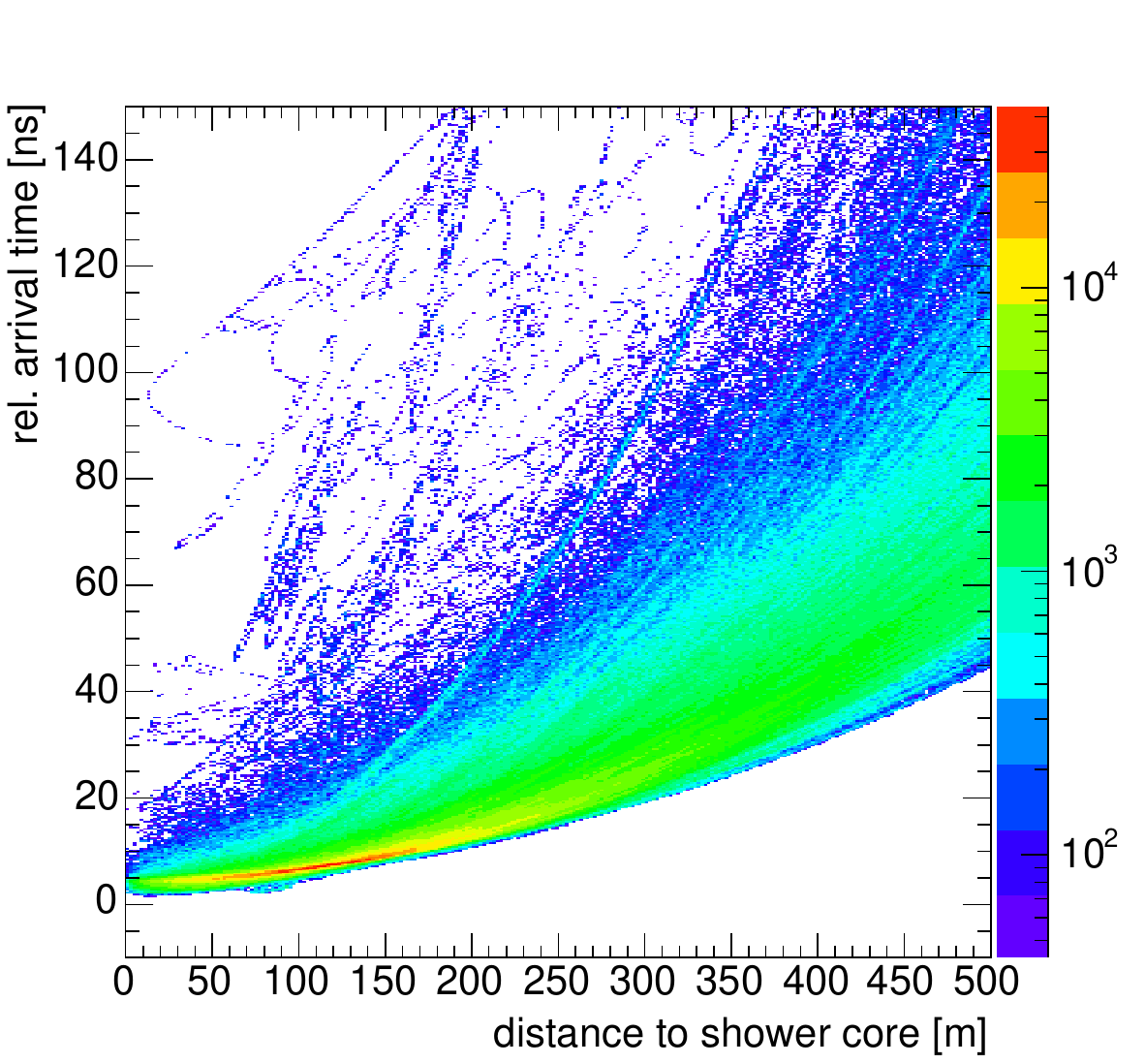}
\caption{{\it Left:} Air Shower Front of a gamma-ray-like event from Crab Nebula. The horizontal axis denotes the distance from the shower core, while the vertical axis denotes the residual of the plane fit to the actual particle detection time of each air shower array (HAWC) counter. It shows that the air shower front is not a plane but a curved surface, and the time spread is several nanoseconds within the radius of $\sim$ 100~m. Original figure is from \cite{Abeysekara_2017}. {\it Right:} Arrival time distribution of Cherenkov photons on the ground as a function of radial distance to the shower core for a vertical incidence 300\,GeV photon shower. Observatory altitude is set at 1,800\,m above sea level. The curvature of the shower front with a width of a few nanoseconds can be distinguished. Simulations were performed using the \texttt{CORSIKA} package \cite{1998cmcc.book.....H}. Courtesy of Gernot Maier \cite{gernot_maier_2022_6037688}.}
 \label{f:gamma_time}
\end{figure}

While the geometry of a full electromagnetic shower resembles a cone (Fig.~\ref{f:showers}), its actual time-resolved development is, at first order, a down-going disk centered at the shower axis and moving at the speed of light. In fact, the shower front has 
a concave shape similar to that of a contact lens although thinner in the center and thicker toward the edges. This geometry of the shower front is due to particles at the edges of the shower having longer travel times as well as a broader arrival time distribution due to multiple Coulomb scattering. Near the core of the shower, the thickness of the shower front is $\lesssim 10$\,ns (Fig.~\ref{f:gamma_time}). The density distribution of electrons and positrons ($\rho_\textrm{e}$) as a function of the radial  distance $r$  from the shower core can be parameterized using the Nishimura-Kamata-Greisen lateral  distribution function \cite{1960ARNPS..10...63G}
\begin{equation}
\label{eq:nkg}
    \rho_\mathrm{e} (r,s,x)= \dfrac{N(x)}{R_\mathrm{M}^2} \ \dfrac{\Gamma(4.5-s)}{2\pi\,\Gamma(s)\,\Gamma(4.5-2s)}\ 
    \left(\dfrac{r}{R_\mathrm{M}}\right)^{s-2} 
    \left(1+\dfrac{r}{R_\mathrm{M}}\right)^{s-4.5}
\end{equation}
valid for shower ages ranging $0.5 \leq s \leq 1.5$. 

The Earth's magnetic field has a second-order effect on the development of electromagnetic showers, as the directions of electrons and positrons are deflected in opposite directions by the geomagnetic field \cite{2008NIMPA.595..572C,2013APh....45....1S}. This effect is non-negligible when compared to multiple Coulomb scattering, and its relative importance increases for low-energy showers. In addition, the Lorentz force systematically deflects particles in opposite directions depending on the sign of their charge, while the effects of multiple Compton scattering are random. At first order, geomagnetic fields stretch the lateral distribution of shower secondaries, reducing the particle density on the ground and making detection more difficult. The Earth's magnetic field also introduces systematic differences that depend on the azimuthal direction of the shower axis for non-vertical showers that complicate the reconstruction of shower parameters from experimental ground-based data.

\subsection{Cherenkov light}
\label{sect:cherenkov}
Charged particles in a 
medium with refractive index $n$ moving with speed $v > \, c/n$ will emit Cherenkov radiation. Cherenkov light is observed in underwater nuclear reactors and has recently been detected in the vitreous humor of patients undergoing radiation therapy \cite{TENDLER2020422}.
Cherenkov light produced by charged secondaries  can be used in ground-based detection of gamma rays. 

\begin{figure}[!t]
 \centering
 \includegraphics[height=8cm]{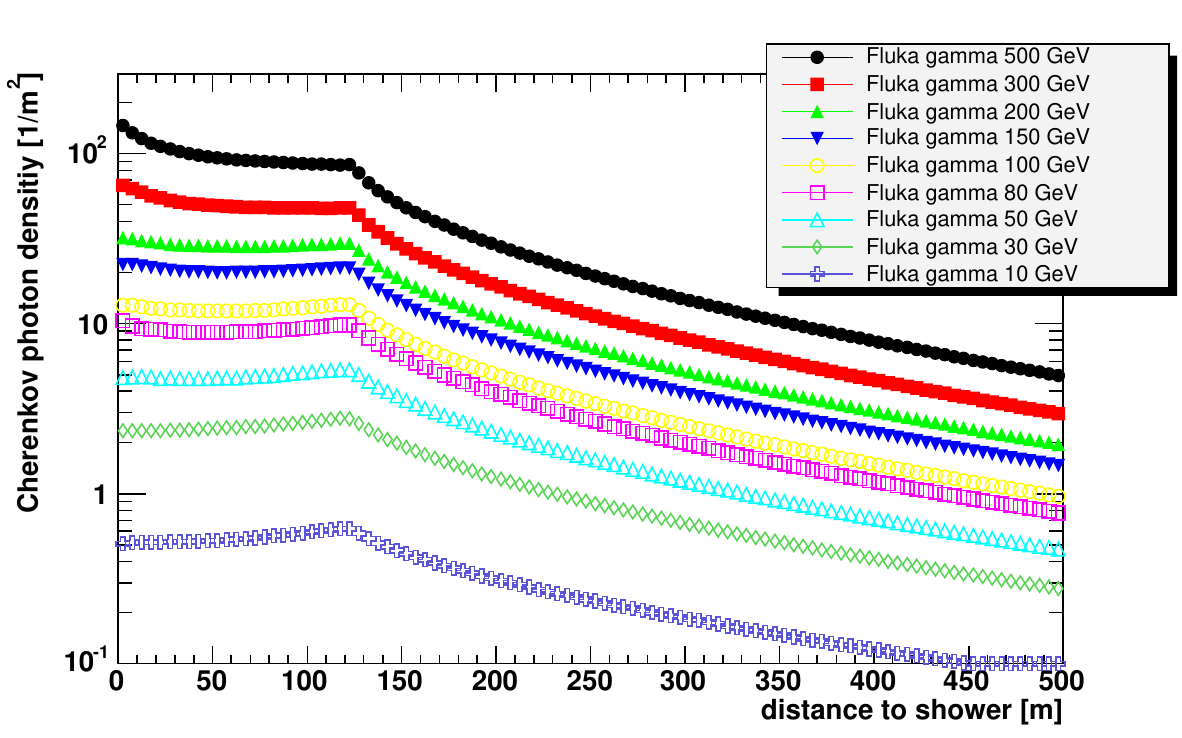}
 \caption{Lateral distribution of Cherenkov photons as a function of radial distance to the shower core, averaged for 1,000 vertical-incidence simulated showers with different primary energies. Observatory altitude is set at 1,800\,m above sea level. A light pool with radius $\sim 130$\,m  can be seen.  Simulations where performed using the \texttt{CORSIKA} package \cite{1998cmcc.book.....H}. Courtesy of Gernot Maier \cite{gernot_maier_2022_6037688}.}
 \label{f:gamma_lat}
\end{figure}

\begin{figure}[!t]
 \centering
 \includegraphics[height=4.5cm]{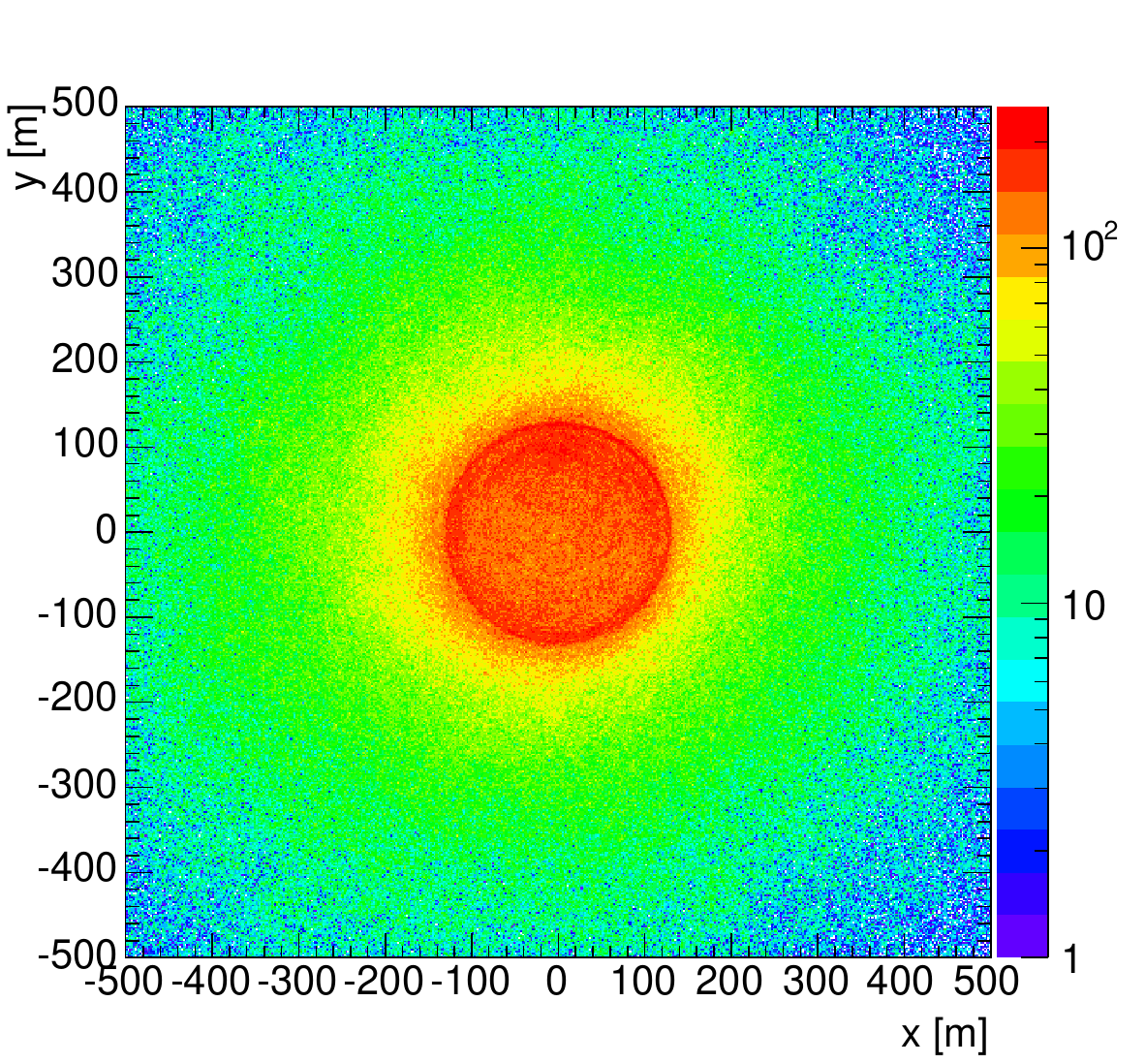}
  \put(-119,103){\colorbox{white}{\color{black} 0.3\,TeV photon}}
 \hspace{1cm}
 \includegraphics[height=4.5cm]{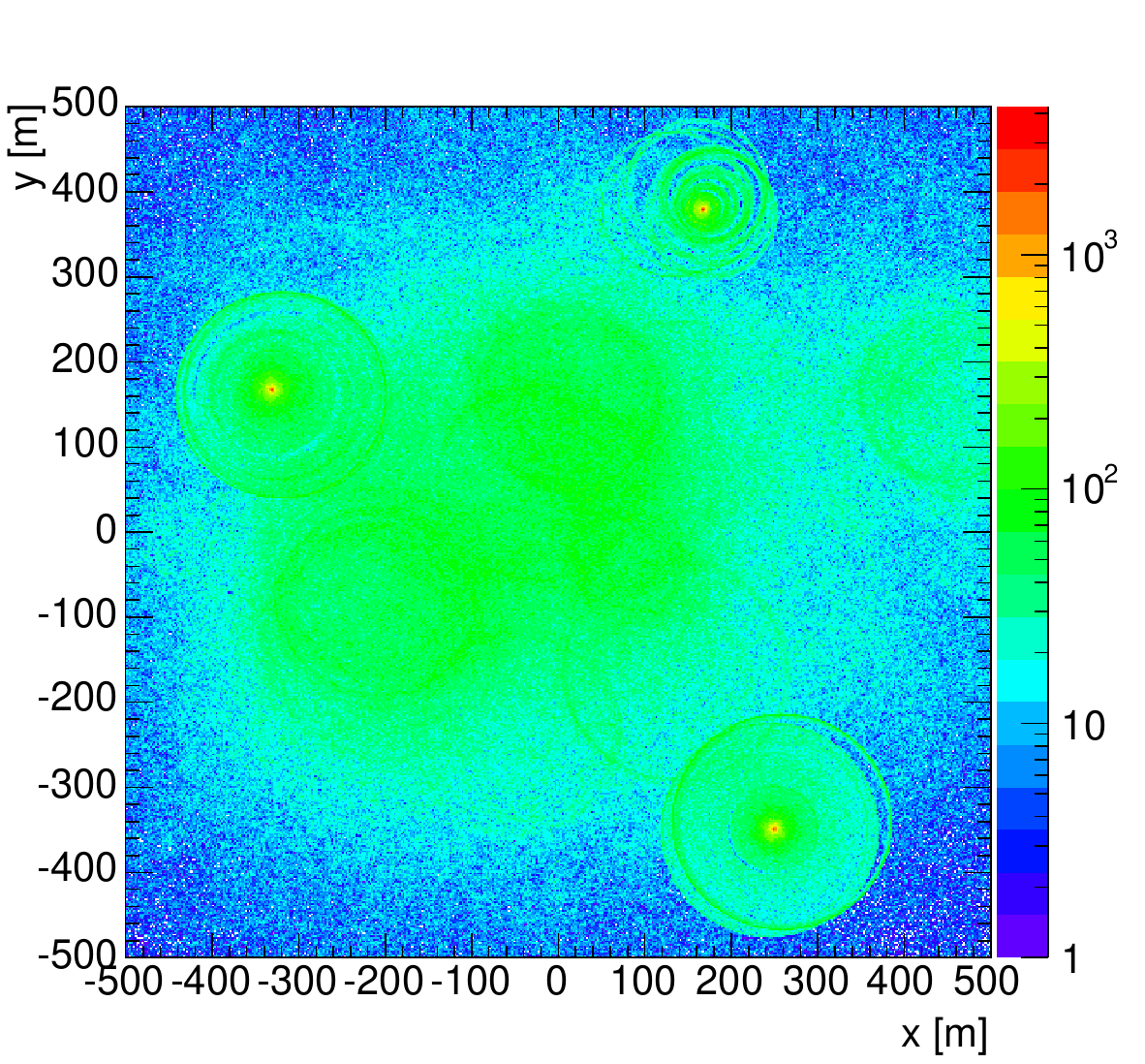}
  \put(-119,103){\colorbox{white}{\color{black} 0.5\,TeV proton}}
 \caption{Density of Cherenkov photons on the ground for a vertical incidence 300\,GeV photon shower and a 500\,GeV proton shower. Observatory altitude is set at 1,800\,m above sea level. The proton shower shows additional clumpyness due to subshowers initiated be the production of pions with large transverse momentum. Simulations were performed using the \texttt{CORSIKA} package \cite{1998cmcc.book.....H}. Courtesy of Gernot Maier \cite{gernot_maier_2022_6037688}.}
 \label{f:gamma_proton}
\end{figure}

Considering the Earth atmosphere as a dielectric medium with refractive index $n(z)$, particles with mass $m$ and energy $E=\gamma\, m c^2$ will produce Cherenkov light if their Lorentz factor is 
\begin{equation}
\label{eq:ch_th}
    \gamma\  \geq \dfrac{n(z)}{\sqrt{n(z)^2-1}}
\end{equation}
The dependence of the atmospheric refractive index with height is a function of the air density and can be approximated by
\begin{equation}
    n(z) = 1.0 + 0.000283\,\dfrac{\rho(z)}{\rho(z=0)}
\end{equation}
Cherenkov radiation is emitted at the Cherenkov angle $\theta_\textrm{Ch}$ such that $\cos{\theta_\textrm{Ch}} = c/v\, n(z)$. At $z=10$\,km 
$\theta_\textrm{Ch} = 12$\,mrad or $0.8^\circ$ and the Cherenkov  energy threshold $E_\textrm{Ch}$ is approximately 40\,MeV for electrons and positrons and 8\,GeV for muons. The geometry of the Cherenkov light emission at 10\,km height would result in a blurry Cherenkov ring with radius $\sim 10\,\textrm{km}\cdot 0.013 = 130$\,m that determines the  size of the light pool for a typical gamma-ray shower (Fig.~\ref{f:gamma_lat} and \ref{f:gamma_proton}). For a 1\,TeV primary, the photon density inside the light pool is $\sim 100\,\textrm{m}^{-2}$.

The paths of charged secondaries follow an angular distribution $\propto \exp\left(\theta/\theta_0\right)$ with respect to the shower axis, where $\theta_0 = 0.83\left(E_\textrm{Ch}\right)^{-0.67}$. Values of $\theta_0$ are typically in the range between $4^\circ$ and $6^\circ$. The combination of the height-dependent Cherenkov angle and the longitudinal development of the shower gives rise to a characteristic lateral distribution of photons on the ground shown in Fig.~\ref{f:gamma_lat}. The Cherenkov light flash has a width of only a few nanoseconds (Fig.~\ref{f:gamma_time}) and can be the brightest source of light in the sky during the short duration of the pulse. 

The total number of Cherenkov photons $N_\textrm{ph}$ produced by secondaries is proportional to $E_\gamma$ and is 
$\sim 100$ photons per square meter for a 1\,TeV shower. The emitted Cherenkov radiation spectrum follows the Franck-Tamm relation \cite{PhysRev.52.378,Frank:1937fk}
\begin{equation}
    \dfrac{\mathrm{d}^2 N_\textrm{ph}}{\mathrm{d}x\,\mathrm{d}\lambda} = \dfrac{2\pi\alpha}{\lambda^2} \sin^2 \left(\theta_\mathrm{Ch}\right)
\end{equation}
which gives the differential number of Cherenkov photons per unit wavelength $\mathrm{d}\lambda$ and path length $\mathrm{d}x$, with $\alpha$ being the fine-structure constant. Cherenkov photons are affected by atmospheric absorption as they propagate toward the ground level. Rayleigh scattering off of particles smaller than the photon wavelength has a $\lambda^{-4}$ dependence and suppresses the propagation of short wavelengths. Mie scattering on particles with sizes comparable to the photon wavelength depends on the aerosol content of the atmosphere, which can present seasonal or transient variability that needs to be corrected \cite{2009A&A...493..721D,2014APh....54...25H}. Ozone $O_3 + \gamma \longrightarrow O_2 + O$ absorption process that filters off photons in the 200\,nm top 315\,nm range. A combination of the $\lambda^{-2}$ dependence of the Cherenkov emission spectrum and ozone absorption leads to a Cherenkov light distribution at ground level that peaks at $\lambda \approx 300-350$\,nm with a detailed spectral shape depending on the height of the observatory, the height of the shower maximum, and the zenith angle of the observation. 

Geographical differences in atmospheric density profiles lead to differences in Cherenkov light density on the ground of up to 60\%. Seasonal variations at mid-latitude sites have an effect on the order of 15--20\% \cite{BERNLOHR2000255}. Finally, scattering by water vapour in clouds limits the use of atmospheric Cherenkov light to cloudless conditions. The introduction of LIDAR cloud monitoring at observatory sites \cite{2016NIMPA.819...60B} allows for determination of the height of the cloud layer. Gamma-ray observations in the presence of high cloud layers located above the typical location of the shower maximum ($\gtrsim 10-12\,$km) have been conducted with an increased systematic uncertainty in the flux and spectral characterization \cite{2017ApJ...836..205A}. 

\subsection{Differences between electromagnetic and cosmic-ray showers}
\label{sec:hadron_showers}
Protons and heavier nuclei also produce air showers in the atmosphere that constitute the main source of background for ground-based gamma-ray observations. Even for strong gamma-ray sources, the signal-to-background shower rates are $10^{-3}-10^{-4}$. Hadronic interactions in cosmic-ray showers lead to the production of secondary nucleons and pions. A characteristic difference with respect to  pair production and bremsstrahlung is the larger transverse momentum carried by hadronic interactions, leading to showers with larger lateral spread (Fig.~\ref{f:showers}). Neutral pions decay quickly into two gamma-ray photons producing an electromagnetic subshower. Charged pions decay into a muon and a neutrino. Muons do not suffer from multiple Coulomb scattering and propagate straight to the ground, producing a characteristic Cherenkov light ring. With a lifetime of 2.2\,ns, a significant fraction of muons reach ground level before decaying due to relativistic time dilation. 
Differences in morphology (Fig.~\ref{f:showers} and \ref{f:gamma_proton}) between electromagnetic and cosmic ray showers, as well as their muon content, constitute the main means of discrimination between gamma-ray-initiated showers and cosmic rays. 

\subsection{Air shower simulations}
Monte Carlo simulations are used in modern particle physics and cosmic ray experiments to understand the development of electromagnetic and hadronic cascades in a complex medium. The study of air showers through simulations has been key to the success of ground-based gamma-ray astronomy. In seminal work from 1985 Hillas exposes that ``it should be possible to distinguish very effectively between background hadronic showers and TeV gamma-ray showers from a point source on the basis of the width, length and orientation of the Cherenkov light images of the shower'' \cite{1985ICRC....3..445H}. These Monte Carlo studies, as well as hardware developments such as the use of pixelated cameras, enabled the development of the imaging Cherenkov technique that lead to the detection of TeV emission from the Crab Nebula \cite{1989ApJ...342..379W} and the blossoming of ground-based gamma-ray astronomy seen in the following decades.  

Over the years, the  \texttt{CORSIKA} software package \cite{1998cmcc.book.....H} has become the field standard for simulations of particle showers in the Earth's atmosphere.  Other standard simulation packages include \texttt{GEANT4} \cite{AGOSTINELLI2003250,ALLISON2016186}, \texttt{EGS} \cite{2000MedPh..27..485K}, and \texttt{FLUKA} \cite{2014snam.conf06005B,2022FrP.....9..705A} and are used to simulate specific aspects of the shower development as well as the response of detectors and telescopes. 

\section{Air shower particle detectors}
\label{sec:sampling} 
Electromagnetic cascades (air showers) initiated by very high-energy gamma rays develop deep in the atmosphere (Fig.~\ref{f:lon_dist}). At sufficient altitude above sea level, a significant number of secondary particles will reach the ground, and can be detected with {\it air shower particle detectors}, also known as particle sampling arrays. 
At 4000\,m height, for example, the typical spread of secondary particles in an air shower is $\sim 100$ m in radius, and they arrive in a few nanoseconds long bunch (see the left panel of Fig.~\ref{f:gamma_time}).  An array of particle detectors deployed in the area of $\sim 10,000-1,000,000$\,m$^2$ can detect these particles and reconstruct the arrival direction and the energy of the primary high-energy gamma rays. 

One method to detect air shower particles is to use scintillators. Arrays of plastic scintillators with $\sim 1$\,m$^2$ surface and a few centimeters thick can be used to sparsely cover a large surface area. Electrons and positrons with energies down to a few MeV produce scintillation light that can be collected and detected with photomultipliers. Photons carry a significant fraction of the total shower energy, but plastic scintillators only produce a response to charged particles.  To overcome this limitation, it is common to cover the scintillator with $\sim 1$ radiation lengths of lead or another metal with high atomic number to convert the photons to electron-positron pairs without significantly absorbing the electron flux.  
 The Chicago Air Shower Array (CASA), the Tibet Air Shower gamma experiment (Tibet-AS$\gamma$), and the Large High Altitude Air Shower Observatory (LHAASO) are some of observatories that use scintillator arrays. 
Resistive plate counters can also be used as surface particle detectors. 
Resistive plate counters are composed of a thin, gas-filled  detector with two metal plates and two high-resistance ($\sim 10^{10} \rm{\Omega}$ cm) plates set up as a large planar capacitor. A high voltage of about $\sim~10$ kV is applied between the metal plates. When a charged particle passes though the detector, the gas along the particle track is ionized, producing an electrical discharge between the plates. Due to the high-resistivity layer, the avalanche effect is quenched quickly and stays confined in the region along the particle track, providing good time ($\sim 1$ ns, \cite{Aielli2006}) and spatial ($\sim 100 \mu$m, \cite{John2022}) resolution. Resistive plate counters are used in particle collider experiments (e.g. ATLAS). The ARGO-YBJ experiment adopted this technology  for air shower detection, realizing a carpet particle detector over an area of $\sim 6,700$\,m$^2$ that achieves a gamma-ray detection threshold as low as 100\,GeV \cite{Bacci1999}.
 
A third method to detect air shower particles is to use Cherenkov light emission in water. The refractive index of the water is $\sim 1.33$ and electrons and positrons with energy above 0.78\,MeV emit Cherenknov light (see Equation~\ref{eq:ch_th}). This Cherenkov radiation is emitted at a Cherenkov angle of 
A large container of water such as a pool or a lake can be used as a particle sampling array by installing photomultiplier tubes in it to collect Cherenkov light, with every single photomultiplier acting 
as an independent counter. 
The Milagro gamma-ray observatory realized a water Cherenkov array on a 4,800\,m$^2$ pool with an array of photomultipliers placed under 1.3\,m of water with a spacing of $\sim 3$\,m between photomultipliers \cite{2007ApJ...658L..33A}. A smaller array of photomultipliers at 6\,m depth was used to measure the penetrating component of the showers. 
A second way to realize a water Cherenkov array is to deploy isolated 
water tanks instrumented with 
one or more photomultipliers. 
The HAWC observatory is an example of such array with three hundred 7\,m diameter $\times$ 5\,m depth water tanks that are densely packed to cover an area on 22,000\,m$^2$ \cite{Abeysekara_2017}.

 \subsection{Event reconstruction with air shower particle detectors}
The lateral distribution of air shower secondaries peaks at the shower axis and decays exponentially  away from it (see Eq.~\ref{eq:nkg}). The location of the intersection between the shower axis and the detector plane is called the shower core. 
The location of the shower core can be estimated by computing the center of gravity of the particle densities detected across several detector units of the air shower array. 
Moreover, as can be seen in the left panel of Fig.~\ref{f:gamma_time}, the shower front arriving on the ground has a characteristic curved shape. Using the arrival time information 
 of the particles in the shower front on each detector together with the estimated core location, the arrival direction of the primary gamma-ray can be estimated. The precision of estimation depends on the number of detected particles as well as the density of the array and the number of detectors that have a signal above threshold. 

The energy of the primary gamma-ray can be estimated from the number of detected secondary particles that can be derived from the total signal in the detector array. 
In most cases, the detector array is located at a height that lies below the shower maximum. 
Therefore, the conversion from the number of detected particles to the primary gamma-ray energy is not fully deterministic and relies an indirect assumption of the location of the shower maximum that is generally based on comparison of the observed data with Monte Carlo simulations of electromagnetic showers. 

Observatories with densely packed or {carpet} arrays (water detectors or resistive plate counter arrays) have a low energy threshold with sufficient gamma/hadron separation ability, while their  effective area is limited by the surface covered by the array. A more sparse deployment of detectors on the ground improves the effective area at high energies at the cost of also rising the energy threshold of the observatory.

\subsection{Cosmic-ray rejection with air shower particle detectors}
\label{sec:gh}
The flux of charged cosmic rays hitting the Earth's atmosphere is significantly higher than the gamma-ray flux, even for the brightest gamma-ray sources. 
Means to distinguish between cosmic-ray and gamma-ray induced showers 
are essential for observatories to reach sufficient signal-to-noise ratio to detect and study gamma-ray sources. 

There are several differences between electromagnetic and hadronic 
showers. One  key feature is 
the muon content of air showers. 
Muons are  produced in hadronic interactions  mainly via the decay of charged pions. On the other hand, electromagnetic showers 
can produce muon secondaries only via photo-pion production or the infrequent creation of a muon-antimuon pair. Therefore, estimating the number of muons detected in air showers can discriminate 
between hadronic and electromagnetic (photon or electron) primaries. 

From the detection point of view, the behavior of muons and electron/positrons inside the detector medium is significantly different. The largest difference is the probability to cause bremsstrahlung. Below $\sim 500$\,GeV, the 
bremstrahlung cross section is inversely proportional to the square of mass of the particle. Therefore, muon bremstrahlung is suppressed by a factor $(m_\textrm{e}/m_\mu)^2 \approx 40,000$ compared to electrons. 
This means that muons are much more difficult to shield from than electrons. 
The radiation length for electrons in typical soil is about $\sim 20$\,cm. About 7 radiation lengths ($\sim 1.4$\,m) of soil  reduce the energy of a 3\,GeV electron down to 3\,MeV, and then ionization takes away the remaining energy. On the other hand, muons  lose their energy  by ionization. After crossing 1.4\,m of soil a 3\,GeV muon loses only 
a tenth of its energy. Because of this penetrating power of muons, 
detectors installed under soil or metal shielding 
will not be sensitive to the electron and photon component of the shower but will detect the muons in the air shower. 

The CASA-MIA experiment installed scintillation counters 3\,m underground 
as muon detectors, while the Tibet AS$\gamma$ observatory used water Cherenkov counters 2\,m underground as primary mean for cosmic-ray rejection through identifying muons. 

\begin{figure*}[!t]
\centering
\includegraphics[width=0.8\hsize]{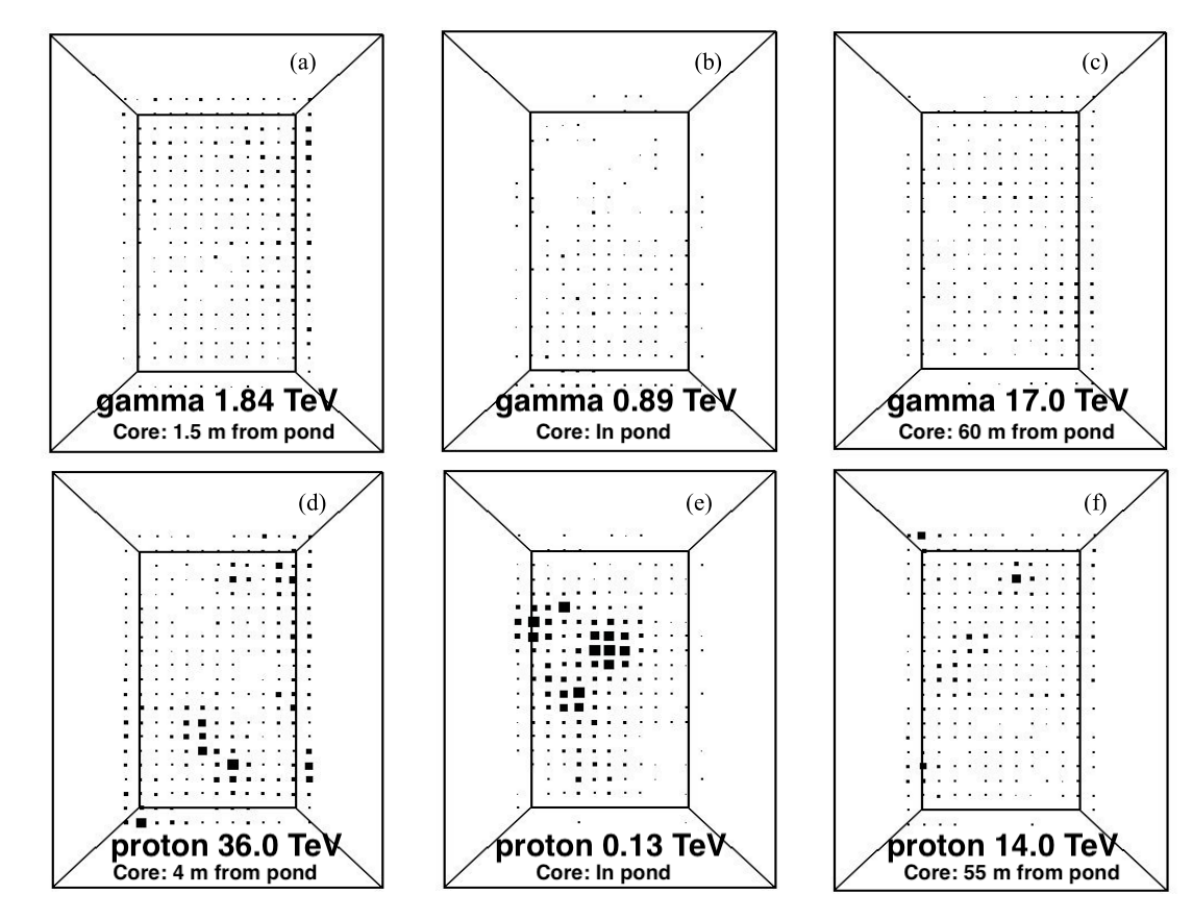}
\caption{Simulated air showers seen by Milagro observatory. 
Top panels are gamma-ray events and bottom panels are proton events. A clear difference in irregularity is seen between gamma rays and protons. Adapted from \cite{Atkins2003}.}
\label{fig:GHSepByMilagro}
\end{figure*}

In absence of dedicated muon detectors, discrimination of cosmic-ray events  can also be accomplished based on the increased  irregularity of distribution of electromagnetic particles in the air shower front in hadronic showers when compared to electromagnetic showers that present smoother particle distribution. Hadronic showers 
can be considered as the collection of electromagnetic subshowers originating from the decay of neutral pions. 
Consequently, the air shower front is not as regular as in  electromagnetic  showers. 
A dense coverage of particle detectors on the ground is necessary to detect the inhomogeneites than can allow for a morphological discrimination between cosmic-ray showers and gamma-ray candidates. 
Fig.~\ref{fig:GHSepByMilagro} shows examples of simulated proton and photon-induced showers as for the Milagro observatory. A large water pond with dense coverage of photomultipliers 
is suitable for capturing  these discriminating features.

\section{Sampling Cherenkov arrays}
\label{sect:ACCA}
\begin{figure}[!tb]
\centering
\includegraphics[width=0.8\hsize]{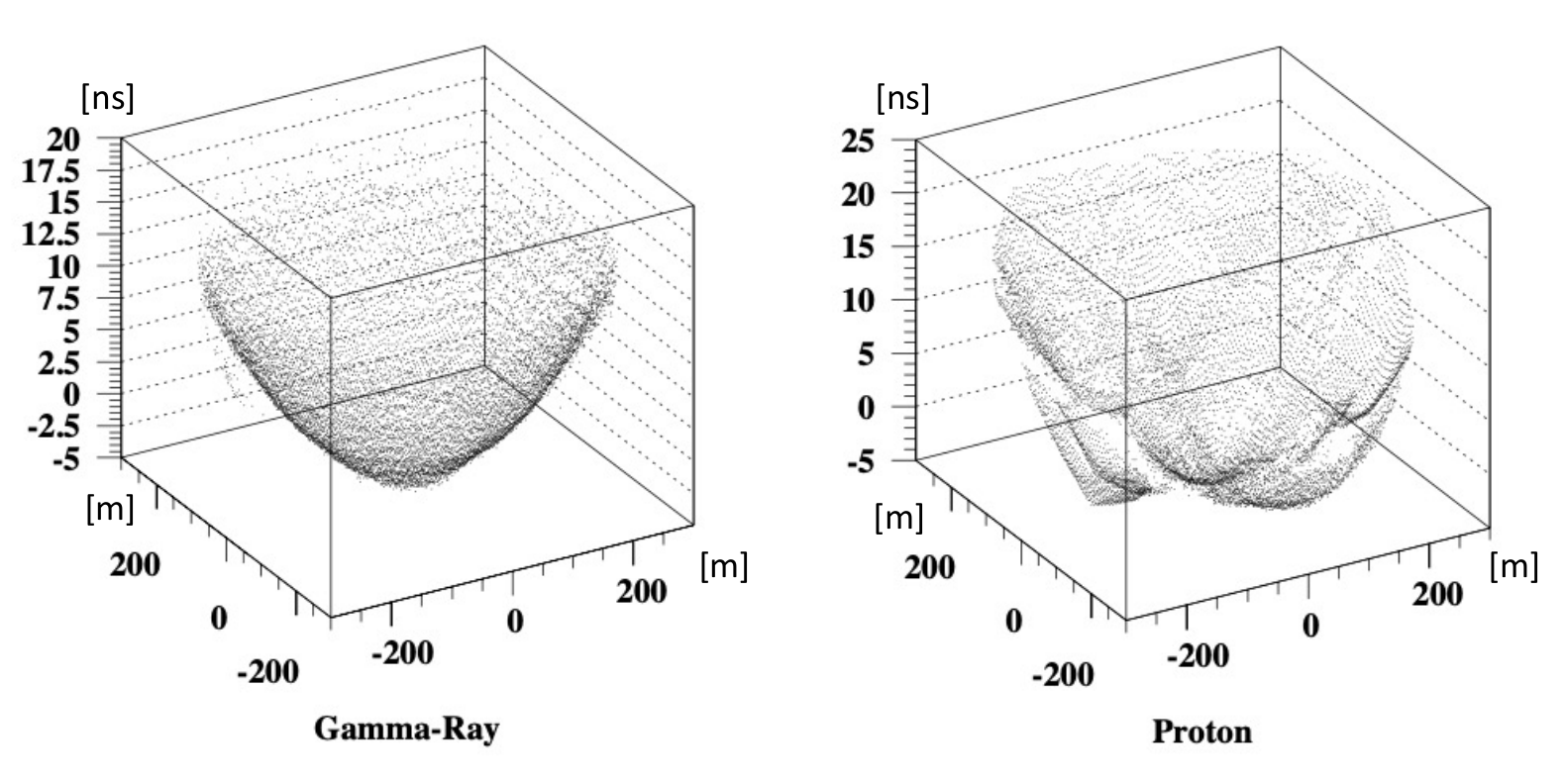}
\caption{Simulation of Cherenkov photon front. X and Y axes denote the position on the ground and the Z axis shows the arrival time. Smoothness is different between a gamma ray (left) and a protons (right).  Adapted from \cite{Oser2001}.}
\label{fig:CherenkovFront}
\end{figure}

Charged particles in air showers emit Cherenkov light.
While UV photons suffer from Rayleigh scattering by aerosols and absorption by ozone molecules, visible light photons reach the Earth's surface where they can be collected and detected using fast photon detectors. 
Cherenkov photons produced during the development of the air shower travel together with the particle shower front, keeping a time spread of a few nanoseconds when they reach ground level. Air showers are seen from the Earth's surface as brief, nanosecond-long light flashes that are 
easy to detect against 
the weak and continuous light background from the night sky. An example of the spatial and time delay distribution of a simulated Cherenkov light shower front from air showers initiated by a gamma-ray photon and a proton 
is shown in Fig.~\ref{fig:CherenkovFront}.

\textit{Sampling Cherenkov arrays} use photomultiplier tubes or mirrors deployed on the ground to cover a large fraction of the $\sim 130$\,m radius Cherenkov light pool of gamma-ray showers.
This can be accomplished either by using bare upward-looking photomultipliers,  short focal length mirrors viewed by individual photon sensors, or with long focal length mirrors each focused onto a photomultiplier in a central camera. In both cases, each mirror is seen by a single photomultiplier that records the light intensity and timing at each position.  
 
Distributed sampling Cherenkov arrays have been used in air shower experiments such as Tunka \cite{2017NIMPA.845..330B}, Yakutsk \cite{1986NIMPA.248..224D}, EAS-TOP \cite{1995APh.....3....1A} or AIROBICC \cite{AIROBICC1996}.  
The AIROBICC detectors in the HEGRA experiment 
consisted of 97 large photomultipliers were deployed in a $\sim 200 \times 200\,{\rm m}^2$ area. Each photomultiplier had a 20\,cm diameter and was pointing toward zenith to directly detect  
Cherenkov photons from air showers. In photomultiplier arrays, shower events can be distinguished from fluctuations of the night sky background by requiring a coincident signal over a number of neighboring detectors within a time window of the order of the duration of the Cherenkov shower light flash. 

\begin{figure*}[!tb]
\centering
\includegraphics[width=0.5\hsize, angle=270]{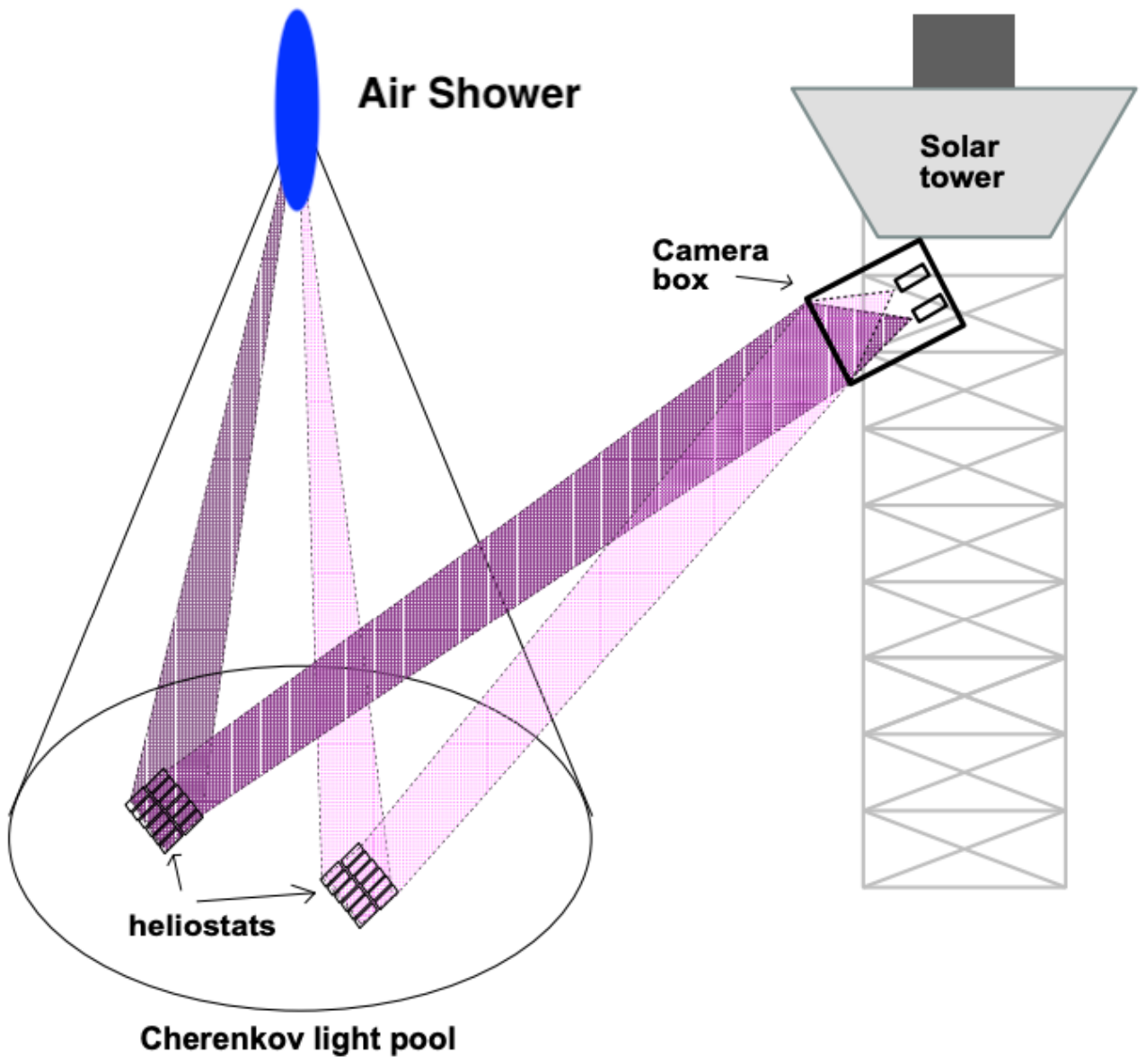}
\caption{A concept of STACEE and CELESTE, adapted from \cite{Ong1996}. Cherenkov light illuminates the large area of the ground and the mirrors distributed over a large area  focus the light to the camera on the tower. Each mirror focuses the light to different photosensors, enabling imaging of the air shower front.}
\label{fig:STACEE}
\end{figure*}

A second implementation of sampling Cherenkov arrays is to use heliostat arrays. 
STACEE \cite{Ong1996} and CELESTE \cite{Smith1997} are two examples of this technique 
(Fig.~\ref{fig:STACEE}), which among other similar experiments played a pioneering role in exploring the energy domain below 100\,GeV. 
Both observatories made use of existing mirror arrays and infrastructure at solar power stations. When operating as a power plant, individual mirrors deployed on the ground are actively operated to track the Sun reflecting its light onto a single heat collector located on a tower at a certain height above the ground. 
At night, when the sun is below the horizon and the solar power plant is not in operation as such, the individual mirrors of the heliostat array can be used to collect Cherenkov photons from air showers and focus them on a photomultiplier camera located on the solar tower. The pointing direction of each mirror is actively adjusted so that Cherenkov photons from a given arrival direction falling on each heliostat are focused onto a specific photomultiplier in the camera. The output signal of photomultipliers can be used to reconstruct the timing and the light density of Cherenkov photons on the corresponding heliostat, mapping the Cherenkov light pool of the shower on the ground. In other words, the distribution of Cherenkov photons hitting the ground from a given direction 
 over a $\sim 10,000$\,m$^2$ area is projected onto the $\sim 1$\,m$^2$ scale photomultiplier camera. 

\subsection{Event reconstruction and cosmic-ray rejection  with sampling Cherenkov arrays}
Sampling Cherenkov arrays use the positional and arrival time information of Cherenkov photons to reconstruct the main properties of gamma-ray showers. After reconstructing the core position, the arrival direction of the primary gamma-ray is estimated using the reconstructed light intensity and arrival time distribution on the ground as inferred from the pulse height and timing of signals from each individual photomultiplier. 
The shower direction can be geometrically determined from the individual timing of the photomultiplier signals, applying the appropriate corrections due to the curvature of 
the Cherenkov photon front (Fig.~\ref{f:gamma_time}). 
Numerical simulations show that the  density of photons inside the 130\,m-radius Cherenkov light pool is directly proportional to the energy of the primary gamma-ray (Fig~\ref{f:gamma_lat}). Outside of the light pool, the photon density decays with a slope that is related to the atmospheric depth of the shower maximum \cite{1982JPhG....8.1475H}. Local inhomogeneities in the timing and intensity of the signals across the Cherenkov light pool (Fig. \ref{fig:CherenkovFront}) can be used to identify and reject showers initiated by cosmic rays. 

\section{Imaging atmospheric Cherenkov telescopes}

\textit{Imaging air Cherenkov telescopes} (IACTs) use focusing mirror optics and photomultiplier cameras to exploit the angular arrival direction of Cherenkov photons and form images of air showers. 

\begin{figure}[!t]
\centering
\includegraphics[width=0.5\hsize]{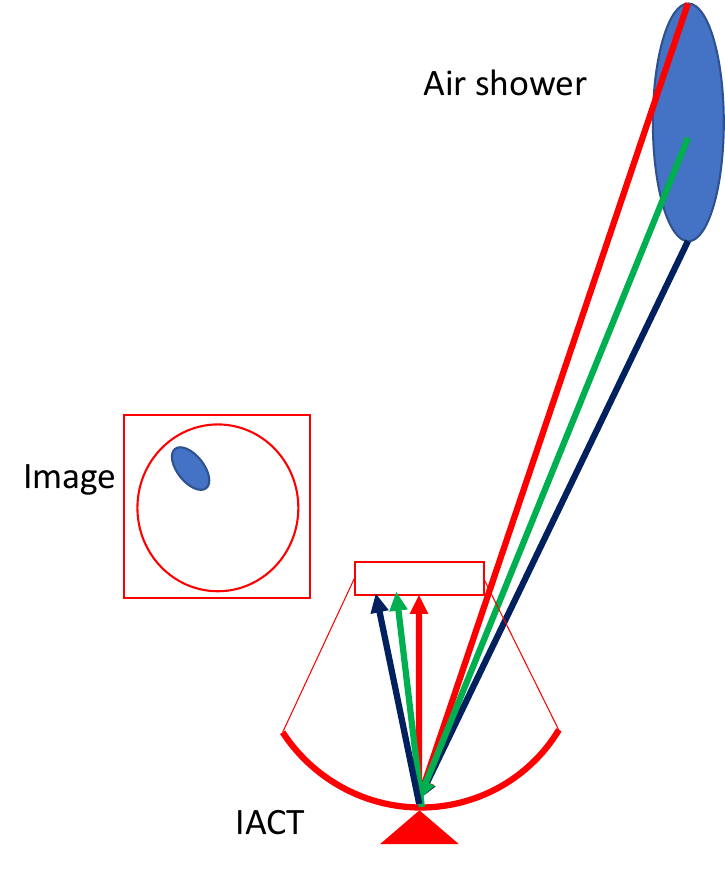}
\caption{Schematic view of Imaging Atmospheric Cherenkov Telescope technique. The Cherenkov photons emitted at a different altitudes are focused on the different part of the camera by the telescope optics. With this technique, the air shower development can be imaged. }
\label{fig:IACT}
\end{figure}

An imaging Cherenkov telescope is composed of a wide-field optical telescope with a fast photomultiplier camera on its focal plane (Fig.~\ref{fig:IACT}). 
The typical optical assembly consists of a short focal ratio\footnote{focal length $f$ divided by the aperture $D$.} ($f/D\sim 1$) parabolic or Davies-Cotton \cite{1957SoEn....1...16D} reflector constructed with smaller individual mirror segments. The total mirror surface per telescope is typically $\gtrsim 100\,\textrm{m}^{2}$ (up to $>600\,\textrm{m}^{2}$ \cite{2017A&A...600A..89H}) to collect enough photons from a Cherenkov light pool with photon densities of $\sim 10-100\,\textrm{m}^{-2}$ (Fig.~\ref{f:gamma_lat}). 
Alternative approaches to a single-mirror optical design have been tested in recent years. Two-mirror designs based on Ritchey-Chr\'{e}tien optics \cite{1927LAstr..41..541R} can be optimized to cancel aberrations and reduce the plate-scale of the camera resulting on a flat focal plane. Dual mirror Schwarzschild-Couder telescopes have been implemented and commissioned in preparation for the future Cherenkov Telescope Array observatory \cite{2020A&A...634A..22L, 2021APh...12802562A}. 

Focal plane instrumentation consists of an array of photosensors that captures a Cherenkov image of each shower. Single-anode photomultiplier tubes are typically used, providing reasonable photon detection efficiency ($\gtrsim 20$\%), clean amplification ($\sim 10^5$) and nanosecond response. Silicon devices feature in some modern Cherenkov telescope designs \cite{2015arXiv150202582D,2022JATIS...8a4007A}. To contain shower images with impact parameter\footnote{The distance between the shower axis projected on the ground and the telescope location.} out to $\sim130$\,m the camera field of view (FoV) has to be $\gtrsim 3^\circ$ in diameter, requiring hundreds to a few thousand pixels. The angular size of a 100\,GeV shower seen at zenith is $\sim0.1^\circ$. Showers with higher energy primaries, observed at higher zenith angles or with large impact parameters have larger angular extents. To resolve the shower structure, it is important that the optical angular resolution of the telescope, the angular size of the camera pixels, and the pointing accuracy of the telescope should all be $\lesssim 0.1^\circ$.

A single Cherenkov telescope located inside the shower light pool can record and potentially reconstruct the energy and the arrival direction of the primary gamma-ray from the shower image. Therefore, the effective area of an IACT is determined by the size of the Cherenkov light pool, which is of the order of $5 \times 10^5\,\textrm{m}^2$.  
To record Cherenkov images, IACTs are triggered when a number $m$ of pixels (typically 2-4) in the camera exceed a certain discrimination threshold (in photoelectrons) within a narrow time window $\tau$ which is typically a few nanoseconds long. This reduces the rate of false triggers from fluctuations of the night sky background $R_\textrm{NSB}$ well below the individual pixel rate $R_1$ following $R_\textrm{NSB} \sim R_1^m \tau^{m-1}$. Most IACTs use topological requirements such as forcing the triggering pixels to be adjacent or located in the same camera sector to further reduce the accidental trigger rate. 

Most observatories consist of more than one imaging Cherenkov telescope, making use of the stereoscopic imaging technique. Having multiple views of a given shower from different perspectives enhances the geometrical reconstruction of the shower. In addition, requiring showers to trigger in more than one camera further reduces the accidental trigger rate due to the night sky background as well as from local muons that at high impact parameters produce gamma-like shower images. The sensitivity of an IACT array increases roughly as the square root of the number of telescopes. The optimal spacing depends on the energy rage to be covered, with wider spacing improving the sensitivity at higher energies. Finally, the performance of the array improves when the array becomes significantly larger than the Cherenkov light pool, achieving better sensitivities for low energy showers even with telescopes of modest size 
\cite{2009APh....32..221C,2013APh....43..171B}.

\subsection{Event Reconstruction and cosmic-ray rejection with IACTs}

The optics of an imaging Cherenkov telescope convert photon angular arrival directions into 
positional information in the focal plane camera (Fig.~\ref{fig:IACT}). Electromagnetic air showers produce elliptical shower images when recorded by an imaging Cherenkov telescope. The longitudinal development of the shower is mapped onto the major axis of the elliptical shower image, with the image being offset from the arrival direction of the shower primary by an angular distance proportional to the shower impact parameter. 
In a stereoscopic system, the intersection of the major axes of the image ellipses obtained by different telescopes indicates the arrival direction of the primary particle. This results in an energy-dependent angular resolution that typically reconstructs 68\% of the gamma rays from a point source within $\sim 0.1^\circ$ of the source position. 

The Cherenkov photon yield of a shower is proportional to the number of secondaries which is in turn proportional to the energy of the primary particle. The amount of light collected in shower images is a good indicator of the primary gamma-ray energy, although it also depends on the impact parameter of the shower. Stereoscopic energy reconstruction dramatically improves the energy resolution of IACTs, with images from multiple telescopes allowing to break the degeneracy between shower energy and impact parameter. This leads to energy resolutions of $\sim 15$\% above 1\,TeV.

Showers induced by cosmic rays are the dominant background for ground-based detection of gamma rays. Event selection based on the \textit{width} and \textit{length} of shower images can efficiently reject cosmic-ray showers while keeping most gamma-ray events \cite{1985ICRC....3..445H}. These cuts exploit the more compact nature of gamma-ray showers, which have a smaller lateral spread than hadronic showers. 
Cosmic-ray showers are essentially a superposition of multiple electromagnetic cascades originating from the decay of individual neutral pions. These irregularities in the shower development propagate to the shower images and can also be 
exploited for cosmic ray rejection. Gamma-ray-looking cosmic ray showers that survive background rejection cuts will have an isotropic distribution of arrival directions. Searches for point sources of gamma rays use the clustering of the arrival directions of gamma-ray events around the source location as one final means to extract the gamma-ray signal from the background. 

\section{Complementarity between ground-based techniques}

Typical performance parameters for a selection of ground-based gamma-ray observatories
are summarized in Table \ref{tab:Performance}.
Air shower arrays and IACTs have complementary capabilities. Combined coverage of sky regions with both techniques provides sensitivity to detect and study a wide variety of source classes. Some of the main differences between techniques are operational: 
IACTs run as pointed telescopes and cover a limited $\lesssim 5^\circ$ region of the sky at any given time while air shower arrays are sensitive to gamma-ray sources located directly overhead of the observatory within a $\sim 2$\,sr cone. In addition, IACTs are essentially optical telescopes and can only operate during clear nights without significant moonlight, leading to duty cycles 
$< 15\%$. Air shower arrays do not suffer from this limitation and have duty cycles well in excess of 90\%.
At the highest energies, where event statistics limits the sensitivity of the observatory and the product of integration time and effective area dominates the capability to detect steady sources, air shower arrays have an advantage over IACTs. Air shower arrays are also well suited to study  extended ($\gtrsim1^\circ$) sources, bright transient objects, and to produce unbiased sky surveys that lead to source population studies and serendipitous discoveries. 
Current-generation IACTs have the best instantaneous point-source sensitivity thanks to their low energy threshold and have superior angular resolution, energy resolution, and background rejection capabilities compared to air shower arrays. In the energy range above 10\,TeV the sensitivity of detectors based on the Cherenkov technique is limited by the effective area of the observatory. Wavefront sampling experiments can deploy small and widely separated detectors and achieve very large effective areas in a more efficient way than regular IACTs. 

\setlength{\tabcolsep}{4pt} 
\begin{threeparttable}[tb]
\centering
\caption{Main properties of selected list of ground-based gamma-ray observatories.}
    \begin{tabular}{clrrrrrr}
\hline \hline
\makecell{obs. \\ type}  & name  & alt. & FoV   & \makecell{duty \\ cycle}   & \makecell{ang.\\ res.\tnote{a}}   &\makecell{energy \\ thresh.} &  \makecell{point source \\sensitivity\tnote{b}}\\ 
 &        & \myalign{c}{[km]} & \myalign{c}{[$^\circ$]} &         & \myalign{c}{[$^\circ$]}        & \myalign{c}{[TeV]}    & \myalign{c}{[Crab Unit]}\\ \hline
  \multirow{3}*{\makecell{particle \\ sampling}} 
  & Tibet AS$\gamma\tnote{c}$  &4.3  & $\sim 45$ & $> 90\%$  & 0.3  & 3    & 20\% \\
  & HAWC   \tnote{d}           & 4.1 & $\sim 45$ & $> 90\%$  & 0.2 & 0.5  & 6\% \\ 
  & \small{{LHAASO}-WCDA}\tnote{e}          & 4.4 & $\sim 45$ & $> 90\%$  & 0.4  & 0.1   & 10\% \\ 
  & \small{{LHAASO}-KM2A}\tnote{e}          & 4.4 & $\sim 40$ & $> 90\%$  & 0.4  & 30   & 1\% \\ \hline
  \multirow{3}*{\makecell{Cherenkov \\ sampling}} 
  & AIROBICC\tnote{f}          & 2.2 &  $\sim 45$& $\sim 10\%$ & 0.3        & 15    & $\sim 100 \%$\\ 
  & CELESTE\tnote{g}           & 1.6 & $\sim 1$  & $\sim 10\%$ & $\sim 0.5$ & 0.03  & 12\%\\ 
  & STACEE\tnote{h}            & 1.7 & $\sim 1$  & $\sim 10\%$ & $\sim 0.5$ & 0.1   & 40\% \\ \hline
  \multirow{3}*{\makecell{Cherenkov \\ imaging}} 
  & H.E.S.S.\tnote{i}         & 1.8  & 5.0 & $\sim 12\%$& 0.05 & 0.02 & 0.7\%\\ 
  & MAGIC\tnote{j}            & 2.2  & 3.5 & $\sim 12\%$& 0.08 & 0.03 & 0.7\%\\ 
  & VERITAS\tnote{k}          & 1.3  & 3.5 & $\sim 10\%$& 0.09 & 0.15 & 0.7\%\\ \hline\hline
    \end{tabular}
    \label{tab:Performance}
    \begin{tablenotes}[para,flushleft,online,normal] 
\item[a] {\footnotesize Values reported at 1\,TeV except for air shower arrays and AIROBICC, for which the angular resolution is reported at 30\,TeV.} 
\item[b] {\footnotesize The value at the best energy threshold is reported. Sensitivity assumes 1 year of operation for air shower arrays and 50\,h of dedicated exposure for Cherenkov telescopes. }
\item[c] {\footnotesize Ref. \cite{Amenomori_2009}.}
\item[d] {\footnotesize Refs. \cite{ABEYSEKARA2012641,Abeysekara_2017,ABEYSEKARA201326}.}
\item[e] {\footnotesize Ref. \cite{LHAASO}.}
\item[f] {\footnotesize Refs. \cite{AIROBICC1996,AIROBICC2}.}
\item[g] {\footnotesize Refs. \cite{PARE200271,CELESTE2}.}
\item[h] {\footnotesize Refs. \cite{STACEE1,STACEE2,STACEE3}.}
\item[i] {\footnotesize Refs. \cite{HESSCrab,HESSAngle, 2018A&A...620A..66H}.}
\item[j] {\footnotesize Refs. \cite{ALEKSIC201676,MAGICGeminga}.}
\item[k] {\footnotesize Ref. \cite{VERITAS}.}
\end{tablenotes}
\end{threeparttable}
\vspace{0.5cm}

The number of Cherenkov photons in a shower is proportional to the energy of the primary $E_\gamma$. Large-aperture mirrors are required for IACTs to collect enough light to trigger on and reconstruct showers with $E_\gamma \lesssim 100$\,GeV. Smaller Cherenkov telescopes are cheaper to produce, simpler to operate, and could be deployed in large numbers over a larger surface area resulting on a bigger effective area at high energies. A combination of imaging Cherenkov telescopes with different mirror apertures to cover a wide energy range with good sensitivity is the design basis of the future Cherenkov Telescope Array \cite{2013APh....43..171B}. The H.E.S.S. collaboration is currently operating $4\times 12$\,m telescopes surrounding a 28\,m telescope in what constitutes the first hybrid Cherenkov instrument of this kind. The system triggers on events detected either only by the large-aperture telescope (mono) or by any combination of two or more telescopes (stereo) \cite{2017A&A...600A..89H}. 

The combined use different detection techniques in a single observatory has also been explored. The HEGRA array combined up to five 3.3\,m-aperture imaging Cherenkov telescopes with an array of more than $200 \times 1\,\textrm{m}^2$ scintillator counters \cite{HEGRAScint}, a matrix of $77 \times 1$\,sr open photomultipliers acting as a sampling Cherenkov array (AIROBICC), and $17 \times 16\,\textrm{m}^2$ lead-concrete Geiger tracking calorimeters \cite{1996NIMPA.378..399R}. Detectors were operated in a truly hybrid mode, with all systems except for the calorimeter towers contributing to the trigger. For example, AIROBICC would require a signal in six counters to self-trigger but also received external triggers when the imaging Cherenkov telescope array detected a shower \cite{AIROBICC1996}. The Geiger towers would measure the total energy deposition of particles in the shower tail and reconstruct and identify muon and electron tracks.  
The LHAASO observatory is designed on similar hybrid operation principles with a sparse 1.3\,km$^2$ array of scintillator detectors and muon detectors, three large pools that act as water Cherenkov detectors, and 18 wide-field Cherenkov telescopes \cite{LHAASO}. 

\section{Other detection concepts}
A number of alternative  techniques have been explored for detection of gamma rays from the ground. An excellent survey of detection concepts with a historical perspective can be found in \cite{2013APh....43...19H}. Several hybrid concepts have been considered to reduce the sensitivity of imaging Cherenkov telescopes to the cosmic ray background, which was one of the main challenges that frustrated early attempts to detect TeV sources. One technique was to install offset photomultipliers that would detect light originating from $\sim 1^\circ$ angular distance from the shower axis \cite{1975ApJ...201...82G}. An excess of off-axis light would indicate the presence of a muons and would cause the event to be rejected. However, cosmic-ray rejection may have been about a factor of two, and exploiting the shape of the Cherenkov images with multi-pixel cameras achieved better results. 

Different implementations of the imaging Cherenkov technique have also been proposed to lower the energy threshold below 100\,GeV. A large effective area could be achieved using a large number of small telescopes ($>100\times 2.5$\,m aperture). The signals from individual telescopes can be delayed and combined before a trigger is generated, with the potential to achieve a low energy threshold and sensitivity equivalent to that of a single 20-50\,m aperture IACT at a fraction of the cost \cite{2005AIPC..745..748F}. Another concept would consist in deploying imaging Cherenkov telescopes at an altitude of $\sim 5\,km$ above sea level. At these altitudes, a 5\,GeV shower produces photon densities $\sim 1\,\textrm{m}^{-2}$ inside the light pool (compare to Fig.~\ref{f:gamma_lat}) enabling a stereoscopic system of 20\,m telescopes to operate with an energy threshold of $\sim5$\,GeV \cite{2001APh....15..335A}. Shower-to-shower fluctuations and the background from electromagnetic showers initiated by cosmic-ray electrons become increasingly important at such low energies. 

Fresnel lenses enable optics with large apertures and wide fields of view with moderate angular resolution. With a lower weight and cost than reflective optics, a system composed of a 3\,m Fresnel lens and a multi-anode photomultiplier camera could have a similar threshold and sensitivity than a 10\,m class IACT \cite{2011ICRC....9..260C}. Another concept based on Fresnel lenses proposes to use 0.5\,m telescopes developed for the PANOSETI project \cite{2020SPIE11454E..3CM}. Each telescope is equipped with a silicon photomultiplier camera and uses a commercial telescope mount for pointing and tracking. Two PANOSETI telescopes have been tested at the Whipple Observatory site and successfully detected air showers from the Crab Nebula that were identified as gamma-rays with energies of 15-50\,TeV using VERITAS \cite{2022SPIE12184E..8BM}. The compact size of the telescopes and use of commercial technologies reduce the cost by two orders of magnitude compared to classical IACTs. A large array of such low-cost telescopes could bring the benefits of the imaging Cherenkov technique (energy resolution, angular resolution, and background rejection) at energies $> 100$\,TeV. 

Alternative techniques have been proposed to improve the sensitivity and lower the energy threshold of particle sampling arrays. One possibility is to use particle tracking detectors such as time projection chambers \cite{1992ihep.book...81L} instead of scintillator detectors. Tracking detectors do not only record the passage of secondary cosmic-ray particles but can reconstruct their path and arrival direction with $<1^\circ$ precision. Standard particle sampling arrays lose their capability to reconstruct the shower core and axis when the number of detected particles falls below a certain threshold. Tracking detector arrays can use the angular correlation between secondary particles and the main shower axis to achieve good arrival direction reconstruction with only $\sim 20$ recorded tracks \cite{HEINTZE198929}. These can be an order of magnitude fewer detected secondaries than needed in particle sampling arrays, lowering the energy threshold of the experiment. Tracking detectors including an iron plate allows them to identify muons, improving background rejection. An array of up to ten argon-methane gas-filled time projection chambers were operational at the HEGRA site \cite{1996NIMPA.369..284B}. 

Hybrid approaches are also being used to extend the energy range of ground-based observatories. The TAIGA experiment combines wide-angle Cherenkov sampling  (TAIGA-HiSCORE) with small-aperture IACTs and electron/muon detectors \cite{2017NIMPA.845..330B}. TAIGA-HiSCORE presents a cost-effective solution to cover a large area and provide sensitivity in the 100\,TeV regime. IACTs, with a narrower field of view, provide additional background rejection and improved angular resolution for showers detected by both systems. Joint operation of Cherenkov sampling telescopes and the muon array improves background rejection using muon tagging, extending the energy range of the observatory beyond 300\,TeV. 

\section{Conclusion}

Since Galbraith and Jelley first detected Cherenkov light associated with extensive air showers in 1952 \citep{1953Natur.171..349G}, there have been significant advancements in experimental techniques for ground-based detection of gamma rays. The imaging atmospheric Cherenkov technique was developed in the 1980s and further enhanced by exploiting stereoscopic reconstruction during the 1990s. In the early 2000s and 2010s, the field experienced substantial growth as imaging atmospheric telescope arrays detected over 200 gamma-ray sources. Additionally, air shower particle detectors in the early 2020s have demonstrated that wide-field observatories can extend the sensitivity of ground-based observatories to energies in the tens of TeVs and beyond.

Currently, there are plans for new and improved observatories, including the hybrid operation of imaging Cherenkov telescopes and particle sampling arrays. These developments, coupled with improved instrumentation and event reconstruction techniques, such as the integration of machine-learning methods to enhance the reconstruction of the arrival direction and energy of primary gamma-rays \citep[e.g.,][]{2018zndo...6842323B,2022NIMPA103966984A}, will ensure ongoing improvements in the sensitivity of future ground-based facilities to cosmic gamma rays.

\bibliographystyle{plain}
\bibliography{main}

\end{document}